\documentclass[prb,aps,twocolumn,showpacs]{revtex4-1}

\usepackage{graphicx,color}
\usepackage{amsthm}
\usepackage{amsfonts}
\usepackage{algorithmic}
\usepackage{enumerate}
\usepackage{latexsym}
\usepackage{amsmath}
\usepackage{amssymb}
\usepackage{bm}
\usepackage[pdftex,plainpages=false,colorlinks=true,linkcolor=blue, citecolor=blue,urlcolor=blue]{hyperref}

\emergencystretch=\maxdimen
\begin{document}
\title{Quasiparticle scattering interference in cuprate superconductors}

\author{Minghuan Zeng}
\thanks{These authors contributed equally to this work}
\author{Xiang Li}
\thanks{These authors contributed equally to this work}
\author{Yongjun Wang}
\author{Shiping Feng}
\thanks{Corresponding author. E-mail: spfeng@bnu.edu.cn}


\affiliation{Department of Physics, Beijing Normal University, Beijing 100875, China}


\begin{abstract}
The quasiparticle scattering interference (QSI) is intimately related to the nature
of the quasiparticle and of its interplay with a variety of electronic orders and
superconductivity. Here starting from the microscopic octet scattering model, the
nature of QSI in cuprate superconductors is studied in the $T$-matrix approach. In
particular, a new method of the inversion of matrix is developed to accurately
derive the $T$-matrix for various kinds of impurities, and then the obtained
$T$-matrix is employed to calculate the local density of states (LDOS). It is shown
that the overall features of the LDOS modulation can be described qualitatively by
taking into account the quasiparticle scattering from a single impurity on the
kinetic-energy-driven homogeneous superconducting-state, where the QSI scattering
wave vectors ${\bf q}_{i}$ and the related QSI peak dispersions are internally
consistent within the octet scattering model. However, the pronounced QSI peaks in
the momentum-space LDOS modulation pattern for a single impurity are smeared heavily
in the case for multiple impurities, and then the momentum-space LDOS modulation for
multiple impurities exhibits a speckle pattern. Moreover, the momentum-space LDOS
modulation for Gaussian-random-distribution of on-site impurity at a relatively large
deviation displays a similar behavior of the LDOS modulation for multiple impurities.
The theory also indicates that the impurity weight linearly increases with the
increase of the impurity-scattering strength for the weak scattering strength and
tends to saturate in the strong scattering strength, while it decreases with the
increase of the impurity-scattering screening length for the short screening-length
and saturates in the long screening-length.
\end{abstract}


\maketitle


\section{Introduction}\label{Introduction}

The quasiparticle is one of the most fundamental and ubiquitous physical observables
in cuprate superconductors, carrying the information about the bosonic glue forming
electron pairs\cite{Carbotte11,Bok16}. This follows from an experimental fact that
the unconventional features of the electronic state, including the exceptionally
high superconducting (SC) transition temperature $T_{\rm c}$, are intimately
connected to the particular characteristics of the low-energy quasiparticle
excitation\cite{Damascelli03,Campuzano04,Fischer07,Yin21}. The parent compounds of
cuprate superconductors are Mott insulators\cite{Bednorz86,Wu87} with the layered
crystal structure, where the layered crystal structure is a stacking of the common
CuO$_{2}$ planes separated by other insulating layers, while the localization of an
electron at each copper atom of the CuO$_{2}$ planes in real-space is caused by the
strong electron correlation\cite{Anderson87}. When a small fraction of these
electrons are removed from the CuO$_{2}$ planes, superconductivity characterized by
the delocalization of the momentum-space electron pairs emerges\cite{Bednorz86,Wu87},
indicating that superconductivity and the related exotic physics in the doped regime
are also dominated by the same strong electron correlation. However, this strong
electron correlation also induces the system to find new way to lower its total
energy\cite{Comin16,Vishik18}. This tendency leads to that the system harbours
diverse manifestations of electronic orders, which coexist or compete with
superconductivity\cite{Comin16,Vishik18}, and induce the electronic inhomogeneity
\cite{Pan01,Vershinin04}. This electronic inhomogeneity is manifested itself as
spatial variations in the local density of states (LDOS). On the other hand, in
addition to the change of the charge-carrier concentration, the doping process
nearly always introduces some measure of disorder\cite{Hussey02,Balatsky06,Alloul09},
therefore in principle, all cuprate superconductors have naturally impurities (or
disorder). More specifically, although the electronic inhomogeneity observed in LDOS
is intrinsic in the nature\cite{Pan01,Vershinin04}, the quasiparticles scattering
from impurities interfere with one another, which leads to that the electronic
inhomogeneity is seeded by this impurity scattering
\cite{Pan00,Hoffman02,McElroy03,Hanaguri07,Kohsaka08,Hanaguri09,Lee09,Vishik09,Schmidt11,Fujita14}.
These experimental observations therefore show that the particular characteristics
of the quasiparticle excitation in cuprate superconductors are dominated by both the
strong electron correlation and impurity scattering.

Experimentally, scanning tunneling microscopy/spectroscopy (STM/S) is the only
measurement technique to probe the real-space inhomogeneous electronic structure of
cuprate superconductors\cite{Fischer07,Yin21}. In particular, STM/S has been used
to infer the momentum-space information of the quasiparticle from the
Fourier-transform of the position (${\bf r}$)- and energy ($\omega$)-dependent LDOS,
then both the real-space and momentum-space LDOS modulations are explored
simultaneously\cite{Fischer07,Yin21}. The LDOS modulation results in the real-space
quasiparticle standing waves, characterized by the wave vectors ${\bf q}$ connecting
points on the constant energy contour in momentum-space. The typical feature in the
momentum-space LDOS modulation is dominated by the peaks. From these peaks as a
function of energy, one can obtain the energy variation of the momentum ${\bf q}$,
which is an autocorrelation between the electronic bands $E_{\bf k}$ and
$E_{{\bf k}+{\bf q}}$. For the same band $E_{\bf k}$, the intensity distribution of
the LDOS modulation can be different, depending on the quasiparticle scattering
geometry. However, this quasiparticle scattering geometry is closely correlated
with the shape of the constant energy contour and the related electronic structure,
which are directly associated with the single-particle excitation spectrum
\cite{Damascelli03,Campuzano04}. In other words, the LDOS modulation intensity is
proportional to the intensities of the single-particle excitation spectra at the
momenta ${\bf k}$ and ${\bf k}+{\bf q}$, while the peak in the modulation then are
corresponding to the highest joint density of states\cite{Fischer07,Yin21}. After
intensive investigations over more than three decades, a large body of data
available from the STM/S measurement technique have provided rather detailed
information of the LDOS modulation in cuprate superconductors
\cite{Hussey02,Balatsky06,Alloul09,Pan00,Hoffman02,McElroy03,Hanaguri07,Kohsaka08,Hanaguri09,Lee09,Vishik09,Schmidt11,Fujita14},
where some essential agreements have emerged: (i) the most part of the LDOS
modulation appeared in the SC-state can be described by the quasiparticle scattering
interference (QSI) caused by the quasiparticle scattering from impurities, where
the QSI peaks are located at the well-defined scattering wave vectors ${\bf q}_{i}$
obeying an {\it octet scattering model}; (ii) all the QSI peaks in the LDOS
modulation have appreciable dispersion; and (iii) the evolution of the dispersions
with doping is consistent with the change of the constant energy contour. These
essential agreements therefore provide the firm evidence for the coexistence of QSI
and superconductivity.

Although a number of the important consequences from the microscopic inhomogeneities
together with the associated fluctuating ordering phenomena have been well-explored
experimentally
\cite{Hussey02,Balatsky06,Alloul09,Pan00,Hoffman02,McElroy03,Hanaguri07,Kohsaka08,Hanaguri09,Lee09,Vishik09,Schmidt11,Fujita14},
the full understanding of the nature of the inhomogeneity is still a challenging
issue. Theoretically, the homogeneous part of the SC-state electron propagator in
the preceding discussions is based on the modified Bardeen-Cooper-Schrieffer (BCS)
formalism with the d-wave symmetry\cite{Hussey02,Balatsky06,Alloul09}, and then
LDOS is derived in terms of the perturbation theory in the Born limit
\cite{Capriotti03,Wulindan10,Nowadnick12,Torre16,Torre16NP}, the self-consistent
$T$-matrix approach\cite{Wang03,Zhang03,Zhu04a,Wangshuhua15,ZhaoMM23},
the Bogoliubov-de Gennes equations\cite{Kreisel15,Choubey17}, and the recently
developed method starting directly from the real-space electron propagator
\cite{Sulangi17,Sulangi18-a,Sulangi18-b}, for a single impurity or a finite impurity
concentration.
The strong electron correlation plays a crucial role in the inhomogeneity
\cite{Comin16,Vishik18}, since the inhomogeneity is intrinsic to the electronic
structure\cite{Pan01,Vershinin04}. However, some of these calculations
\cite{Capriotti03,Wulindan10,Nowadnick12,Torre16,Torre16NP,Wang03,Zhang03,Zhu04a,Wangshuhua15,ZhaoMM23,Kreisel15,Choubey17,Sulangi17,Sulangi18-a,Sulangi18-b},
(i) suffer from ignoring the
strong electron correlation in the homogeneous part of the phenomenological BCS-type
electron propagator, or include the partial strong electron correlation effect by
adding a phenomenological electron self-energy into the homogeneous part of the
BCS-type electron propagator\cite{Wulindan10}, or consider the partial strong
electron correlation effect in terms of the renormalization mean-filed treatment of
the homogeneous part of the BCS-type electron propagator\cite{Wangshuhua15}, and
therefore the homogeneous part of the phenomenological electron propagator in these
calculations
\cite{Capriotti03,Wulindan10,Nowadnick12,Torre16,Torre16NP,Wang03,Zhang03,Zhu04a,Wangshuhua15,ZhaoMM23,Kreisel15,Choubey17,Sulangi17,Sulangi18-a,Sulangi18-b}
can not give a consistent description of the electronic structure observed by the
angle-resolved photoemission spectroscopy (ARPES) experiments
\cite{Damascelli03,Campuzano04,Fischer07};
(ii) the ARPES experiments\cite{Chatterjee06,McElroy06,Chatterjee07,Restrepo23,He14}
have shown clearly that
the Fermi arcs that emerge due to the electron Fermi surface (EFS) reconstruction
at the case of zero energy\cite{He14,Norman98,Shi08,Sassa11,Comin14,Kaminski15}
can persist into the case for a finite binding-energy, where a particularly large
fraction of the spectral weight is located at around the tips of the Fermi arcs.
These tips of the Fermi arcs connected by the scattering wave vectors ${\bf q}_{i}$
thus construct an {\it octet scattering model}, which can give a consistent
description of the regions of the highest joint density of states observed in the
ARPES autocorrelation experiments
\cite{Chatterjee06,McElroy06,Chatterjee07,Restrepo23,He14}. On the other hand, the
STM/S experiments have also demonstrated unambiguously that the distinguished
features of QSI are dominated by the peaks at the scattering wave vectors
${\bf q}_{i}$ obeying an {\it octet scattering model}
\cite{Hussey02,Balatsky06,Alloul09,Pan00,Hoffman02,McElroy03,Hanaguri07,Kohsaka08,Hanaguri09,Lee09,Vishik09,Schmidt11,Fujita14}.
In this case, two crucial requirements in theoretically reproducing results from
STM/S experiments are the homogeneous part of the strongly correlated electron
propagator and the microscopic octet scattering model. However, to the best of our
knowledge, the nature of QSI in cuprate superconductors has not been discussed
starting from a microscopic SC theory, and no explicit calculations of LDOS based
on the microscopic octet scattering model has been made so far.

In the recent study\cite{Zeng22}, we have discussed the influence of the impurity
scattering on the electronic structure of cuprate superconductors, where the
homogeneous part of the electron propagator is obtained based on the
kinetic-energy-driven SC mechanism, and then this electron propagator produces the
shape of the constant energy contour and the related microscopic octet scattering
model as measured by the ARPES experiments
\cite{Chatterjee06,McElroy06,Chatterjee07,Restrepo23,He14}, while the
impurity-scattering self-energy is derived in terms of the self-consistent
$T$-matrix approach. The obtained results\cite{Zeng22} also show that the decisive
role played by the impurity-scattering self-energy in the particle-hole channel is
the further renormalization of the quasiparticle band structure with a reduced
quasiparticle lifetime, while the impurity-scattering self-energy in the
particle-particle channel induces a strong deviation from the d-wave behaviour of
the SC gap, indicating that the highly unconventional features of the electronic
structure are generated by both the strong electron correlation and impurity
scattering. In this paper, we study respectively the quasiparticle scattering
from a single impurity and multiple impurities along with this line. Starting from
the homogeneous part of the electron propagator and the related microscopic octet
scattering model, we develop a new method of the inversion of matrix to accurately
derive the $T$-matrix for various kinds of impurities, where the calculation of the
$T$-matrix contains all the quasiparticle excitations and scattering processes.
The obtained $T$-matrix then is employed to evaluate LDOS, where we respectively
examine in details the LDOS modulations for an in-plane single impurity, an
out-of-plane single impurity, in-plane multiple impurities, out-of-plane multiple
impurities, and Gaussian-random-distribution of on-site impurity. Our results show
that although there are some subtle differences between the LDOS modulations for an
in-plane single impurity and an out-of-plane single impurity, the overall features
of the LDOS modulation in cuprate superconductors can be reproduced qualitatively
by taking into account the quasiparticle scattering from a single impurity on the
kinetic-energy-driven homogeneous SC-state, where the QSI peaks appear at the
scattering wave vectors ${\bf q}_{i}$, while these QSI wave vectors ${\bf q}_{i}$
and the related QSI peak dispersions can be consistently explained in terms of the
octet scattering model. However, the pronounced QSI peaks in the momentum-space
LDOS modulation pattern for a single impurity are broadened in the case of
multiple impurities, and then the outcome of the LDOS modulation for multiple
impurities is almost a speckled version of the single-impurity result. Moreover,
the momentum-space LDOS modulation for Gaussian-random-distribution of on-site
impurity at a relatively large deviation displays a similar behavior of the LDOS
modulation for multiple impurities.

The rest of this paper is organized as follows. The methodology is presented in
Sec. \ref{Formalism}, while the quantitative characteristics of the LDOS modulation
are presented in Section \ref{Results}, where we show that the impurity weight
increases with the increase of the impurity-scattering strength for the weak
scattering strength, while it decreases with the increase of the impurity-scattering
screening length for short screening length.
Moreover, the essential features of the LDOS modulation remain unchanged even in
the presence of the filter effect except for that the filter effect induces a local
d-wave symmetry around the parallel direction of the LDOS phase in momentum-space.
Finally, we give a summary in
Sec. \ref{conclude}. In Appendix \ref{matrix}, we present the accurate derivation
of the impurity-induced $T$-matrix in terms of the new method of the
inversion of matrix.

\section{Methodology}\label{Formalism}

\subsection{Model and strong electron correlation}
\label{model-constraint}

The only common characteristic in the layered crystal structure of cuprate
superconductors is the presence of CuO$_{2}$ planes\cite{Bednorz86,Wu87}, and it
seems evident that the intrinsic features, including the unconventional
superconductivity, in cuprate superconductors are governed by the CuO$_{2}$ plane
\cite{Carbotte11,Bok16,Damascelli03,Campuzano04,Fischer07,Yin21}. In this case, it
has been proposed that the essential physics of the doped CuO$_{2}$ plane can be
properly accounted by the single-band $t$-$J$ model on a square lattice
\cite{Anderson87,Anderson87a,Zhang88,Lee06,Edegger07,Spalek22},
\begin{eqnarray}\label{tjham}
H&=&-t\sum_{l\hat{\eta}\sigma}C^{\dagger}_{l\sigma}C_{l+\hat{\eta}\sigma}
+t'\sum_{l\hat{\eta}'\sigma}C^{\dagger}_{l\sigma}C_{l+\hat{\eta}'\sigma}
+\mu\sum_{l\sigma} C^{\dagger}_{l\sigma}C_{l\sigma}\nonumber\\
&+&J\sum_{l\hat{\eta}}{\bf S}_{l}\cdot {\bf S}_{l+\hat{\eta}},
\end{eqnarray}
supplemented by an important on-site local constraint
$\sum_{\sigma}C^{\dagger}_{l\sigma}C_{l\sigma}\leq 1$ to remove the double electron
occupancy, where the summation is over all sites $l$, and for each $l$, over its
nearest-neighbors (NN) sites $\hat{\eta}$ or the next NN sites $\hat{\eta}'$,
$C^{\dagger}_{l\sigma}$ and $C_{l\sigma}$ are the electron operators that
respectively create and annihilate an electron with spin orientation
$\sigma=\uparrow,\downarrow$ on lattice site $l$, ${\bf S}_{l}$ is spin operator
with its components $S^{\rm x}_{l}$, $S^{\rm y}_{l}$, and $S^{\rm z}_{l}$, and
$\mu$ is the chemical potential.
It should be emphasized that this single-band $t$-$J$ model (\ref{tjham}) has
been derived from a multi-band Hubbard model described the CuO$_{2}$ plane
\cite{Zhang88}, and then the unconventional features of cuprate superconductors
and the related superconductivity\cite{Zhang88,Lee06,Edegger07,Spalek22} have been
discussed intensively based on the $t$-$J$ model (\ref{tjham}). More specifically,
the ARPES experimental observations\cite{Damascelli03,Campuzano04,Fischer07}
indicate that the single-band $t$-$J$ model (\ref{tjham}) is of particular
relevance to the low energy features of cuprate superconductors.
Throughout this paper, the NN magnetic exchange
coupling $J$ and lattice constant $a$ are set as the energy and length units,
respectively, while the NN hopping amplitude $t$ and next NN hopping amplitude $t'$
are set to be $t/J=2.5$ and $t'/t=0.3$, respectively, as in our previous discussions
\cite{Zeng22}.

To incorporate the on-site local constraint of no double electron occupancy in the
$t$-$J$ model (\ref{tjham}), the fermion-spin transformation has been developed
\cite{Feng0494,Feng15}, where a spin-up annihilation and a spin-down annihilation
operators for the physical electrons are represented as the composite operators,
\begin{eqnarray}\label{css}
C_{l\uparrow}=h^{\dagger}_{l\uparrow}S^{-}_{l}, ~~~
C_{l\downarrow}=h^{\dagger}_{l\downarrow}S^{+}_{l},
\end{eqnarray}
respectively, and then the motion of the constrained electrons are restricted in
the restricted Hilbert space without double electron occupancy in actual analyses,
where the $U(1)$ gauge invariant spinful fermion operator
$h^{\dagger}_{l\sigma}=e^{i\Phi_{l\sigma}}h^{\dagger}_{l}$
($h_{l\sigma}=e^{-i\Phi_{l\sigma}}h_{l}$) creates (annihilates) a charge carrier on
site $l$, and thus keeps track of the charge degree of freedom of the constrained
electron together with some effects of spin configuration rearrangements due to the
presence of the doped charge carrier itself, while the $U(1)$ gauge invariant spin
operator $S^{+}_{l}$ ($S^{-}_{l}$) keeps track of the spin degree of freedom of the
constrained electron, and therefore the collective mode from this spin degree of
freedom of the constrained electron can be interpreted as the spin excitation
responsible for the dynamical spin response of the system.

\subsection{Homogeneous electron propagator}
\label{model-propagator}

For a microscopic description of the SC-state in cuprate superconductors,
the kinetic-energy-driven SC mechanism has been established
\cite{Feng15,Feng0306,Feng12,Feng15a} based on the $t$-$J$ model (\ref{tjham}) in
the fermion-spin representation (\ref{css}), where the interaction between charge
carriers directly from the kinetic energy by the exchange of the spin excitation
generates a d-wave charge-carrier pairing state in the particle-particle channel,
then the d-wave electron pairs originated from this d-wave charge-carrier pairing
state are due to the charge-spin recombination, and their condensation reveals the
d-wave SC-state. This kinetic-energy-driven SC mechanism reveals that (i) the
constrained electron has dual roles, since the glue to hold the constrained electron
pairs together is {\it the spin excitation, the collective mode from the spin degree
of freedom of the constrained electron itself}. In other words, the constrained
electrons simultaneously act to glue and to be glued\cite{Schrieffer95,Xu23a};
(ii) the spin-excitation-mediated electron pairing state in a way is in turn
strongly influenced by the single-particle coherence, and therefore there is a
competition between the single-particle coherence and superconductivity, which leads
to a dome-like shape doping dependence of $T_{\rm c}$. In the case of the lack of
impurity scattering, no translation-symmetry breaking occurs in the homogeneous
system, and then the diagonal and off-diagonal propagators in the SC-state can be
generally expressed as $G({\bf r},{\bf r}',\omega)=G({\bf r}-{\bf r}',\omega)$ and
$\Im^{\dagger}({\bf r},{\bf r}',\omega)=\Im^{\dagger}({\bf r}-{\bf r}',\omega)$,
respectively. In this case, the homogeneous electron propagator of the $t$-$J$ model
(\ref{tjham}) in the fermion-spin representation (\ref{css}) has been derived in the
previous studies\cite{Feng15a}, and can be expressed explicitly in the Nambu
representation as,
\begin{eqnarray}\label{EGF-NR}
\tilde{G}({\bf k},\omega)&=&\left(
\begin{array}{cc}
G({\bf k},\omega), & \Im({\bf k},\omega) \\
\Im^{\dagger}({\bf k},\omega), & -G({\bf k},-\omega)
\end{array}\right)\nonumber\\
&=&{1\over F({\bf k},\omega)}\{[\omega-\Sigma_{0}({\bf k},\omega)]\tau_{0}
+\Sigma_{1}({\bf k},\omega)\tau_{1}\nonumber\\
&+&\Sigma_{2}({\bf k},\omega)\tau_{2}+[\varepsilon_{\bf k}
+\Sigma_{3}({\bf k},\omega)]\tau_{3}\},~~~
\end{eqnarray}
where $\tau_{0}$ is the unit matrix, $\tau_{1}$, $\tau_{2}$, and $\tau_{3}$ are
Pauli matrices, $\varepsilon_{\bf k}=-4t\gamma_{\bf k}+4t'\gamma_{\bf k}'+\mu$ is
the energy dispersion in the tight-binding approximation, with
$\gamma_{\bf k}=({\rm cos}k_{x}+{\rm cos} k_{y})/2$,
$\gamma_{\bf k}'={\rm cos}k_{x}{\rm cos}k_{y}$,
$F({\bf k},\omega)=[\omega-\Sigma_{0}({\bf k},\omega)]^{2}-[\varepsilon_{\bf k}
+\Sigma_{3}({\bf k},\omega)]^{2}-\Sigma^{2}_{1}({\bf k},\omega)
-\Sigma^{2}_{2}({\bf k},\omega)$, and the homogeneous self-energy
$\Sigma_{\rm ph}({\bf k},\omega)$ in the particle-hole channel has been broken up
into $\Sigma_{\rm ph}({\bf k},\omega)=\Sigma_{3}({\bf k},\omega)
+\Sigma_{0}({\bf k},\omega)$, with its symmetric and asymmetric parts
$\Sigma_{3}({\bf k},\omega)$ and $\Sigma_{0}({\bf k},\omega)$, respectively,
while the homogeneous self-energy $\Sigma_{\rm pp}({\bf k},\omega)$ in the
particle-particle channel has been broken up as $\Sigma_{\rm pp}({\bf k},\omega)
=\Sigma_{1}({\bf k},\omega)-i\Sigma_{2}({\bf k},\omega)$, with its real and
imaginary parts $\Sigma_{1}({\bf k},\omega)$ and $\Sigma_{2}({\bf k},\omega)$,
respectively. Moreover, these homogeneous self-energies $\Sigma_{0}({\bf k},\omega)$,
$\Sigma_{1}({\bf k},\omega)$, $\Sigma_{2}({\bf k},\omega)$, and
$\Sigma_{3}({\bf k},\omega)$ have been obtained in the previous works in terms of
the full charge-spin recombination, and are given explicitly in
Ref. \onlinecite{Feng15a}.
\begin{figure}[h!]
\centering
\includegraphics[scale=0.27]{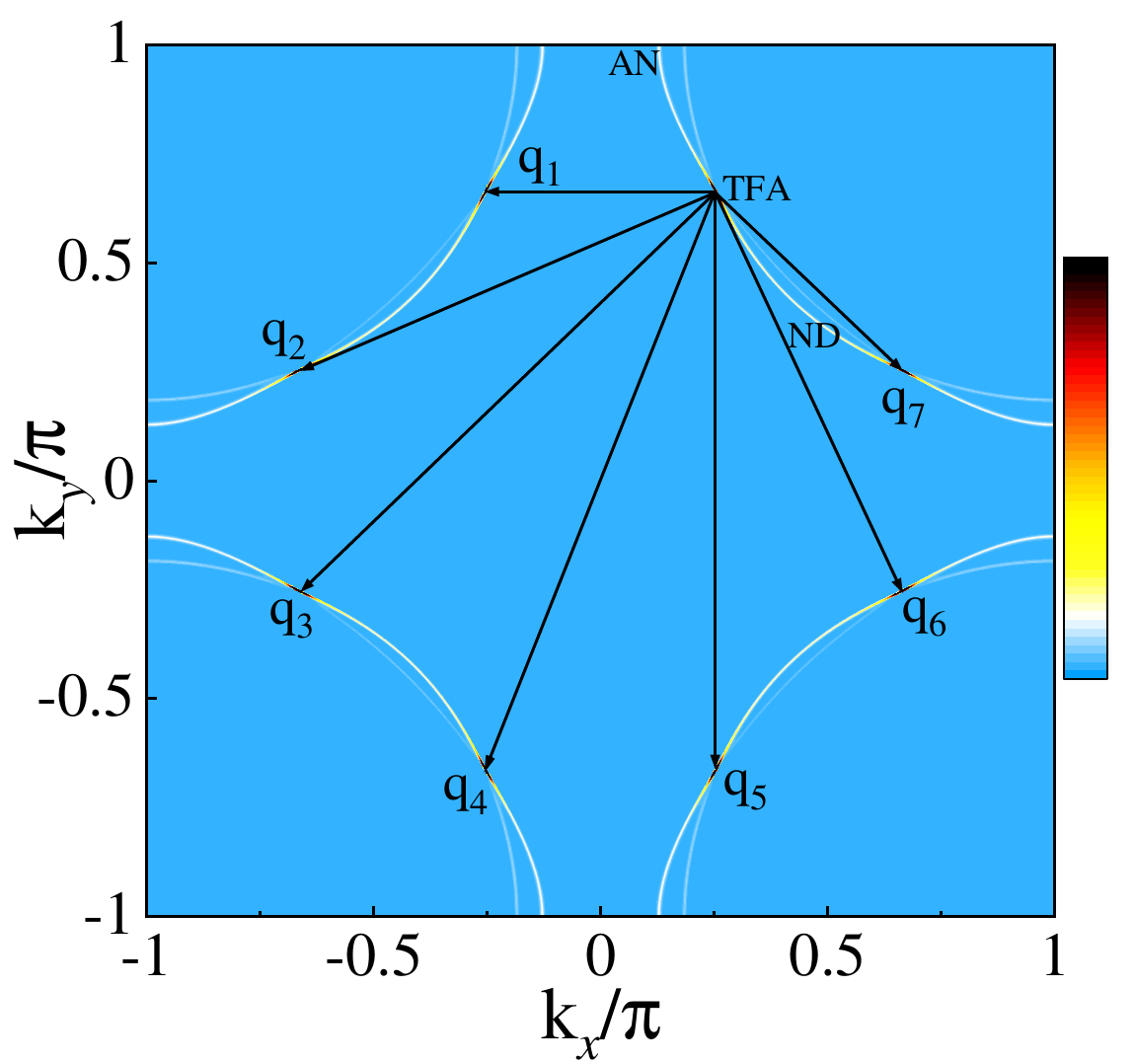}
\caption{(Color online) The intensity map of the homogeneous electron spectral
function for the binding-energy $\omega=-0.16J$ at the doping $\delta=0.15$ with
temperature $T=0.002J$. The tips of the Fermi arcs connected by the scattering wave
vectors ${\bf q}_{1}$, ${\bf q}_{2}$, ${\bf q}_{3}$, ${\bf q}_{4}$, ${\bf q}_{5}$,
${\bf q}_{6}$, and ${\bf q}_{7}$ construct an {\it octet scattering model}.
\label{EFS-map}}
\end{figure}
In particular, the sharp peaks appearing at temperature
$T\rightarrow 0$ in $\Sigma_{0}({\bf k},\omega)$, $\Sigma_{1}({\bf k},\omega)$,
$\Sigma_{2}({\bf k},\omega)$, and $\Sigma_{3}({\bf k},\omega)$ are actually a
$\delta$-function, which are broadened by a small damping used in the numerical
calculation for a finite lattice. The calculation in this paper for
$\Sigma_{0}({\bf k},\omega)$, $\Sigma_{1}({\bf k},\omega)$,
$\Sigma_{2}({\bf k},\omega)$, and $\Sigma_{3}({\bf k},\omega)$ is performed
numerically on a $120\times 120$ lattice in momentum-space, where the infinitesimal
$i0_{+}\rightarrow i\Gamma$ is replaced by a small damping $\Gamma=0.05J$.

Starting from the above homogeneous electron propagator (then the homogeneous
electron spectrum function), the topology of the constant energy contour in the
homogeneous system has been discussed in the previous work\cite{Gao19}. For the
convenience in the following discussions, the intensity map of the homogeneous
electron spectral function
$A({\bf k},\omega)=-{\rm Im}[\tilde{G}(\bm{k},\omega)]_{11}/\pi$ for the
binding-energy $\omega=-0.16$ at the doping $\delta=0.15$ with temperature
$T=0.002J$ is replotted in Fig. \ref{EFS-map}. Apparently, the constant energy
contour is somehow truncated around the first Brillouin zone (BZ) and ends up with
a so-called disconnected Fermi arcs centered around the node
\cite{Chatterjee06,McElroy06,Chatterjee07,Restrepo23,He14}, where we use the
notation {\it Fermi arcs} on the constant energy contour even for a finite binding
energy. However, the electron spectral function (then the density of states)
exhibits the highest intensity at around the tips of the Fermi arcs, and then these
eight tips of the Fermi arcs with the highest density of states connected by the
scattering wave vectors ${\bf q}_{i}$ construct an {\it octet scattering model}.
Within this octet scattering model, the processes from one tip of the Fermi arc to
another become dominant, owing to the highest joint density of states between any
two such points. These dominant scattering processes manifest themselves in a set of
the sharp peaks at seven typical momenta ${\bf q}_{i}$, with $i$=1,2,...7, in the
power spectrum. In particular, the shape of the constant energy contour (then the
octet scattering model) changes with the change of energy, therefore the scattering
wave vectors ${\bf q}_{i}$ should disperse. Moreover, we\cite{Zeng22} have shown
recently that this octet scattering model in the homogeneous system can persist into
the case in the presence of the impurity scattering, although the intensity at the
constant energy contour is reduced. Based on the octet scattering model, the nature
of the ARPES autocorrelation in cuprate superconductors has been investigated
\cite{Zeng22,Gao19}, and the obtained results show that the octet scattering model
shown in Fig. \ref{EFS-map} can give a consistent description of the regions of the
highest joint density of states
\cite{Chatterjee06,McElroy06,Chatterjee07,Restrepo23,He14}.

\subsection{Impurity-dressed electron propagator and the local density of states}
\label{dressed-LDOS}

However, in the presence of impurities with the impurity-scattering potential
$V(\bm{r})$, the homogeneous electron propagator in Eq. (\ref{EGF-NR}) is dressed
by the impurity scattering, where the underlying translation-symmetry of the
square-lattice CuO$_{2}$ plane is broken, and then the impurity-dressed electron
propagator in real-space can be evaluated in terms of the equation of iteration
method as\cite{Hussey02,Balatsky06,Alloul09},
\begin{eqnarray}\label{Green-Fun-Eq}
\tilde{G}_{I}(\bm{r},\bm{r}',\omega)&=&\tilde{G}(\bm{r}-\bm{r}',\omega) \nonumber\\
&+&\sum_{\bm{r}_{1}}\tilde{G}(\bm{r}-\bm{r}_{1},\omega)V(\bm{r}_{1})\tau_{3}
\tilde{G}(\bm{r}_{1}-\bm{r}',\omega) \nonumber\\
&+&\sum_{\bm{r}_{1}\bm{r}_{2}}\tilde{G}(\bm{r}-\bm{r}_{2},\omega)V(\bm{r}_{2})
\tau_{3}\tilde{G}(\bm{r}_{2}-\bm{r}_{1},\omega) \nonumber\\
&\times& V(\bm{r}_{1})\tau_{3}\tilde{G}(\bm{r}_{1}-\bm{r}',\omega)+\cdots.
\end{eqnarray}
With the help of the Fourier transformed impurity-dressed electron propagator,
\begin{eqnarray}
\tilde{G}_{I}(\bm{k},\bm{k}',\omega)={1\over N}\sum_{\bm{r},\bm{r}'}
\tilde{G}_{I}(\bm{r},\bm{r}',\omega)e^{-i\bm{k}\cdot\bm{r}+i\bm{k}'\cdot\bm{r}'},
\end{eqnarray}
and impurity-scattering potential,
\begin{eqnarray}
\sum_{\bm{r}}e^{-i(\bm{k}_{1}-\bm{k}_{2})\cdot\bm{r}}V(\bm{r})=
V(\bm{k}_{1}-\bm{k}_{2})=V_{\bm{k}_{1}\bm{k}_{2}},
\end{eqnarray}
the equation of iteration of the impurity-dressed electron
propagator in Eq. \eqref{Green-Fun-Eq} can be rewritten in momentum-space as,
\begin{eqnarray}\label{Green-Fun-k}
&&\tilde{G}_{I}(\bm{k},\bm{k}',\omega)=\tilde{G}(\bm{k},\omega)
\delta_{\bm{k},\bm{k}'}+\tilde{G}(\bm{k},\omega){1\over N}V_{\bm{k}\bm{k}'}
\tau_{3}\tilde{G}(\bm{k}',\omega)\nonumber\\
&&+{1\over N}\sum_{\bm{k}_{1}}\tilde{G}(\bm{k},\omega)V_{\bm{k}\bm{k}_{1}}
\tau_{3}\tilde{G}(\bm{k}_{1},\omega){1\over N}V_{\bm{k}_{1}\bm{k}'}\tau_{3}
\tilde{G}(\bm{k}',\omega)+\cdots\nonumber\\
&&=\tilde{G}(\bm{k},\omega)\delta_{\bm{k},\bm{k}'}+\tilde{G}(\bm{k},\omega)
\tilde{T}_{\bm{k}\bm{k}'}(\omega)\tilde{G}(\bm{k}',\omega),
\end{eqnarray}
with the $T$-matrix equation\cite{Zeng22},
\begin{eqnarray}\label{T-matrix-equation-1}
&&\tilde{T}_{\bm{k}\bm{k}'}(\omega)={1\over N}V_{\bm{k}\bm{k}'}\tau_{3}
+{1\over N}\sum_{\bm{k}_{1}}V_{\bm{k}\bm{k}_{1}}\tau_{3}
\tilde{G}(\bm{k}_{1},\omega){1\over N}V_{\bm{k}_{1}\bm{k}'}\tau_{3} \nonumber\\
&&+{1\over N}\sum_{\bm{k}_{1}}V_{\bm{k}\bm{k}_{1}}\tau_{3}
\tilde{G}(\bm{k}_{1},\omega){1\over N}\sum_{\bm{k}_{2}}V_{\bm{k}_{1}\bm{k}_{2}}
\tau_{3}\tilde{G}(\bm{k}_{2},\omega){1\over N}V_{\bm{k}_{2}\bm{k}'}\tau_{3}
\nonumber\\
&&+\cdots \nonumber\\\
&&={1\over N}V_{\bm{k}\bm{k}'}\tau_{3}+{1\over N}\sum_{\bm{k}_{2}}
V_{\bm{k}\bm{k}_{2}}\tau_{3}\tilde{G}(\bm{k}_{2},\omega)
\tilde{T}_{\bm{k}_{2}\bm{k}'}(\omega),
\end{eqnarray}
where the contributions to the $T$-matrix can be classified according to how they
modify electronic properties through the impurity scattering potentials: the
conventional impurity scattering occurs in the $\tau_{3}$ channel, while the local
SC gap modification occurs in the $\tau_{1}$ channel. To accurately obtain this
impurity-induced $T$-matrix, we develop a new method of the inversion of matrix
[see Appendix \ref{matrix}], where the $T$-matrix in Eq. (\ref{T-matrix-equation-1})
can be derived explicitly as,
\begin{equation}\label{Tmat-Expression}
\tilde{T}(\omega)=\bar{V}\otimes\tau_{0}{1\over 1-\bar{M}}\hat{I}_{v}\otimes
\tau_{3},
\end{equation}
without any approximations. $\hat{I}_{v}$ is a unit matrix in momentum space, while
the matrices $\bar{V}$ and $\bar{M}$ have been given explicitly in
Appendix \ref{matrix}. In this approach of the inversion of matrix, the calculation
of the $T$-matrix contains all the quasiparticle excitations and scattering
processes.

Substituting the $T$-matrix (\ref{Tmat-Expression}) into Eq. (\ref{Green-Fun-k}),
the impurity-dressed electron propagator $\tilde{G}_{I}(\bm{k},\bm{k}',\omega)$ in
momentum-space can be derived directly, while its expression form in real-space is
obtained by making use of the Fourier-transform as,
\begin{eqnarray}\label{Green-Fun-k-1}
&&\tilde{G}_{I}(\bm{r},\bm{r}',\omega)={1\over N}\sum_{\bm{k},\bm{k}'}
\tilde{G}_{I}(\bm{k},\bm{k}',\omega)e^{i\bm{k}\cdot\bm{r}-i\bm{k}'\cdot\bm{r}'}
\nonumber\\
&&={1\over N}\sum_{\bm{k},\bm{k}'}e^{i\bm{k}\cdot\bm{r}-i\bm{k}'\cdot\bm{r}'}
\tilde{G}(\bm{k},\omega)\delta_{\bm{k},\bm{k}'} \nonumber\\
&&+{1\over N}\sum_{\bm{k},\bm{k}'}e^{i\bm{k}\cdot\bm{r}-i\bm{k}'\cdot\bm{r}'}
\tilde{G}(\bm{k},\omega)\tilde{T}_{\bm{k}\bm{k}'}(\omega)\tilde{G}(\bm{k}',\omega)
\nonumber\\
&&=\tilde{G}(\bm{r}-\bm{r}',\omega)  \nonumber\\
&&+{1\over N}\sum_{\bm{k},\bm{k}'}e^{i\bm{k}\cdot\bm{r}-i\bm{k}'\cdot\bm{r}'}
\tilde{G}(\bm{k},\omega)\tilde{T}_{\bm{k}\bm{k}'}(\omega)\tilde{G}(\bm{k}',\omega).
~~~
\end{eqnarray}
The key quantity of interest in the discussions of QSI is LDOS $\rho(\bm{r},\omega)$
of the system in the presence of impurities. The LDOS modulation
$\delta\rho(\bm{r},\omega)=\rho(\bm{r},\omega)-\rho_{0}(\omega)$ in real-space,
where $\rho_{0}(\omega)$ is the density of states in the case of the absence of
impurities, now can be obtained straightforwardly from the imaginary part of the
diagonal matrix elements of the above impurity-dressed electron propagator
(\ref{Green-Fun-k-1}) as\cite{Hussey02,Balatsky06,Alloul09},
\begin{eqnarray}\label{Rhoq-R}
\delta\rho(\bm{r},\omega)&=& -{2\over\pi}{\rm Im}\Big[{1\over N}\sum_{\bm{k}\bm{k}'}
e^{i(\bm{k}-\bm{k}')\cdot\bm{r}}\tilde{G}(\bm{k},\omega)\nonumber\\
&\times& \tilde{T}_{\bm{k}\bm{k}'}(\omega)\tilde{G}(\bm{k}',\omega)\Big]_{11},
\end{eqnarray}
while its expression form in momentum-space $\delta\rho(\bm{q},\omega)$ can be
evaluated directly via the Fourier-transform as,
\begin{equation}\label{Rhoq}
\delta\rho(\bm{q},\omega)=\text{Re}[\delta\rho(\bm{q},\omega)]
+ i\text{Im}[\delta\rho(\bm{q},\omega)],
\end{equation}
with the real and imaginary parts,
\begin{subequations}
\begin{eqnarray}
\text{Re}[\delta\rho(\bm{q},\omega)]&=&-\frac{1}{\pi}\text{Im}
[\tilde{A}(\bm{q},\omega) + \tilde{A}(-\bm{q},\omega) ]_{11}, \label{Rhoq-Re}\\
\text{Im}[\delta\rho(\bm{q},\omega)] &=&-\frac{1}{\pi}\text{Re}
[\tilde{A}(\bm{q},\omega)- \tilde{A}(-\bm{q},\omega)]_{11},\label{Rhoq-Im}
\end{eqnarray}
\end{subequations}
respectively, where the function $\tilde{A}(\bm{q},\omega)$ can be expressed
explicitly as,
\begin{equation}
\tilde{A}(\bm{q},\omega)=\sum_{\bm{k}}\tilde{G}(\bm{k},\omega)
\tilde{T}(\bm{k},\bm{k}+\bm{q},\omega)\tilde{G}(\bm{k}+\bm{q},\omega).
\end{equation}
However, for a single impurity located at ${\bf r}=0$,
$\tilde{A}(\bm{q},\omega)=\tilde{A}(-\bm{q},\omega)$, which leads to
$\text{Im}[\delta\rho(\bm{q},\omega)]=0$ in Eq. (\ref{Rhoq-Im}), and then the LDOS
modulation in momentum-space
$\delta\rho(\bm{q},\omega)=\text{Re}[\delta\rho(\bm{q},\omega)]$.

The above results in Eqs. (\ref{Rhoq-R}) and (\ref{Rhoq}) therefore show that the
characteristic features of the LDOS modulation are dominated by both the homogeneous
electron propagator $\tilde{G}(\bm{k},\omega)$ in Eq. (\ref{EGF-NR}) [then the
constant energy contour and the related octet scattering model], which reflects the
strong electron correlation effect on the LDOS modulation, and the impurity-induced
$T$-matrix $\tilde{T}(\omega)$ in Eq. (\ref{Tmat-Expression}), which incorporates
both the strong electron correlation and impurity-scattering effects on the LDOS
modulation.

\section{Quantitative characteristics}\label{Results}

The investigation of the characteristic features of QSI can offer insight into the
fundamental aspects of the quasiparticle excitation in cuprate superconductors
\cite{Hussey02,Balatsky06,Alloul09}, and therefore can also offer points of the
reference against which theories may be compared. Since the layered crystal
structure of cuprate superconductors is a stacking of the common CuO$_{2}$ planes
separated by insulating layers as we have mentioned in Sec. \ref{Introduction}, the
impurity distribution accompanied with different types of the doping processes are
quite different, where impurities which substitute for Cu in the CuO$_{2}$ plane
turn out to be strong scatters of the electronic state in the CuO$_{2}$ plane
\cite{Hussey02,Balatsky06,Alloul09}, giving rise to a major modification of the
electronic structure. However, for the cuprate superconductors
(Bi,Pb)$_{2}$(Sr,La)$_{2}$CuO$_{6+\delta}$ and
Bi$_{2}$Sr$_{1.6}$L$_{0.4}$CuO$_{6+\delta}$ (L=La,Nd,Gd), the mismatch in the ionic
radius between Bi and Pb or Sr and L leads to that the effective impurities reside
in the insulating layers\cite{Eisaki04,Fujita05,McElroy05,Kondo07,Hashimoto08} some
distances away from the conducting CuO$_{2}$ plane, where the concentration of the
out-of-plane impurities is controlled by varying the radius of the Pb or L ions.
Moreover, the scattering potential arising from the in-plane impurities is also
different from that arising from the out-of-plane impurities. In this section, we
analyze the quantitative characteristics of QSI generated by various forms of
impurity scattering potentials to shed light on the nature of the SC-state
quasiparticle excitation in cuprate superconductors. In the following discussions,
the strength of the impurity scattering potential is chosen to be positive to avoid
the quantum resonant state, since the quantum resonant state induced by the impurity
scattering potential with the negative strength may leads to that the $T$-matrix
approach breaks down.

\subsection{In-plane single impurity}

In this subsection, we first discuss the distinguishing features of QSI arising
from an in-plane single impurity and of its evolution with doping.
For an in-plane single impurity located at the lattice site ${\bf r}=0$, the
impurity-scattering potential in real-space can be modeled as
\cite{Hussey02,Balatsky06,Alloul09},
\begin{eqnarray}\label{VPOT-RS}
V(\bm{r})={V_{s}\over |\bm{r}|}e^{-|\bm{r}|/L},
\end{eqnarray}
with the impurity-scattering strength $V_{s}$ and screening length $L$, while its
form in momentum-space can be obtained directly in terms of the Fourier-transform
as,
\begin{equation}\label{VPOT-1}
V_{\bm{k}\bm{k}'}={2\pi V_{s}\over \sqrt{(\bm{k}-\bm{k}')^{2}+1/L^2}}.
\end{equation}

\begin{figure*}[t!]
\centering
\includegraphics[scale=0.20]{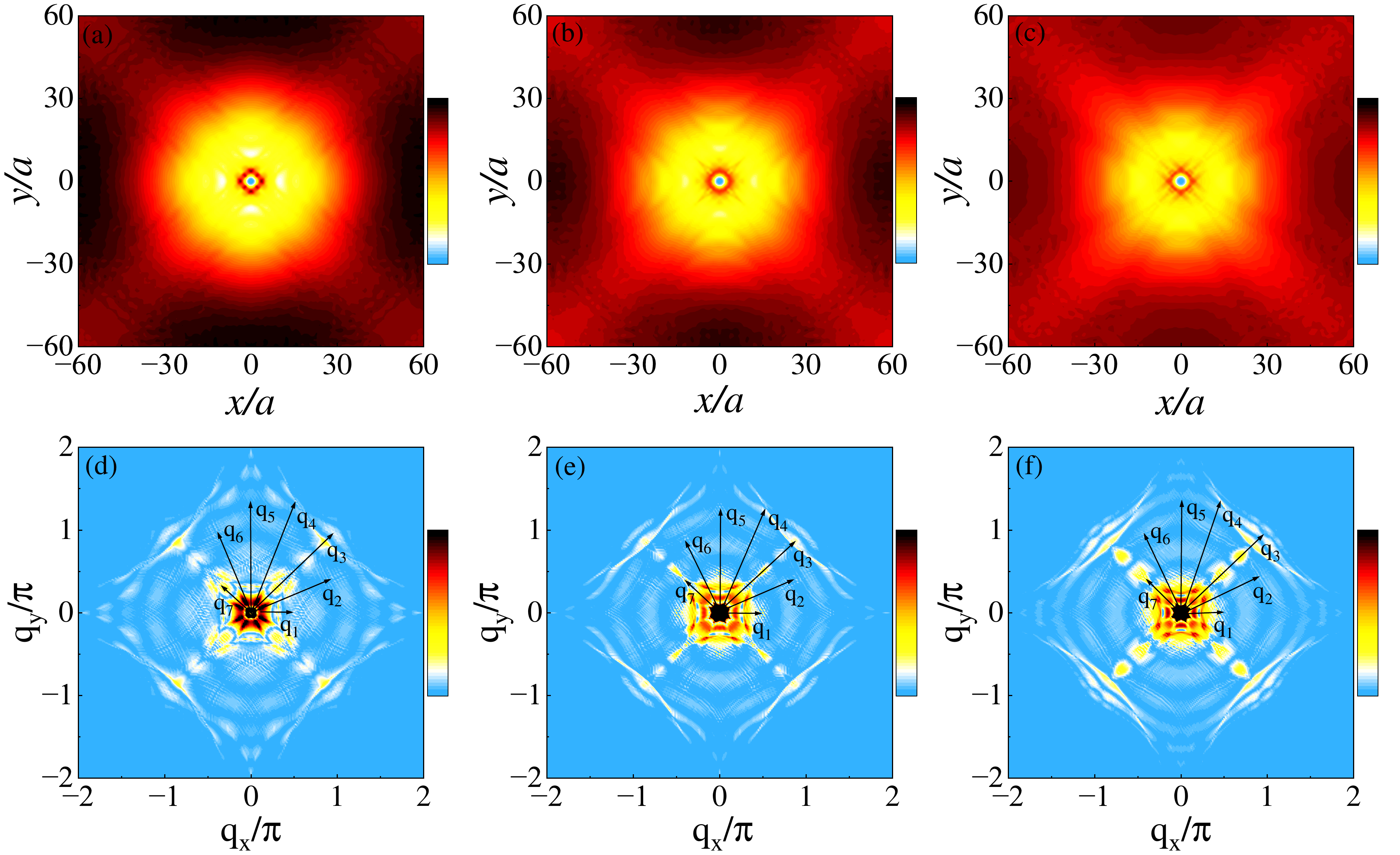}
\caption{(Color online) Upper panel: the maps of the local density of states in
real-space for an in-plane single impurity at (a) $\delta=0.09$, (b) $\delta=0.15$,
and (c) $\delta=0.21$ in $\omega=-0.16J$ and $T=0.002J$ with the impurity-scattering
strength $V_{s}=8J$ and screening length $L=14a$. Lower panel: the corresponding
maps of the local density of states in momentum-space at (d) $\delta=0.09$, (e)
$\delta=0.15$, and (f) $\delta=0.21$ Fourier-transformed from the maps of the
real-space local density of states in (a), (b), and (c), respectively.
\label{LDOS-IPSI}}
\end{figure*}

Substituting the above single impurity-scattering potential (\ref{VPOT-1}) into
Eqs. (\ref{Tmat-Expression}) and (\ref{Green-Fun-k-1}), LDOS in real-space in
Eq. (\ref{Rhoq-R}) and its Fourier transform in momentum-space in Eq. (\ref{Rhoq})
for an in-plane single impurity can be obtained explicitly. In this case, we have
performed a series of calculations for LDOS in real-space and its
Fourier-transformed form with different energies and doping concentrations,
and the results of the LDOS maps in real-space at (a) $\delta=0.09$,
(b) $\delta=0.15$, and (c) $\delta=0.21$ in $\omega=-0.16J$ with $T=0.002J$ for the
impurity-scattering strength $V_{s}=8J$ and screening length $L=14a$ are plotted in
the upper panel of Fig. \ref{LDOS-IPSI}, while the corresponding LDOS maps in
momentum-space at (d) $\delta=0.09$, (e) $\delta=0.15$, and (f) $\delta=0.21$ are
plotted in the lower panel of Fig. \ref{LDOS-IPSI}, which are obtained from the
real-space LDOS maps in (a), (b), and (c), respectively, in terms of the
Fourier-transform.
Obviously, the overall features of the LDOS modulation are similar
to the STM/S observations
\cite{Hussey02,Balatsky06,Alloul09,Pan00,Hoffman02,McElroy03,Hanaguri07,Kohsaka08,Hanaguri09,Lee09,Vishik09,Schmidt11,Fujita14},
and can be summarized as: (i) the spatial pattern exhibits a four-fold ($C_{4}$)
rotational symmetry centered at around the impurity site, which is consistent with
the $C_{4}$ rotation symmetry underlying the square-lattice CuO$_{2}$ plane
\cite{Hussey02,Balatsky06,Alloul09,Pan00,Hoffman02,McElroy03,Hanaguri07,Kohsaka08,Hanaguri09,Lee09,Vishik09,Schmidt11,Fujita14};
(ii) at the impurity site, LDOS is suppressed, and has the smallest intensity.
However, LDOS in real-space does not increase monotonically with distance from
the scatter, but rather oscillates, producing local minima and maxima
\cite{Hussey02,Balatsky06,Alloul09,Pan00,Hoffman02,McElroy03,Hanaguri07,Kohsaka08,Hanaguri09,Lee09,Vishik09,Schmidt11,Fujita14}.
In particular, we have found that different spatial patterns and wave lengths emerge
at each energy; (iii) in accordance with the spatial pattern of LDOS, the LDOS
pattern in momentum-space also has a $C_{4}$ rotational symmetry centered about the
center of BZ. In addition to the brightest spot at around the center of BZ, the
relatively bright discrete-spots appear. In particular, the QSI peaks are located
exactly at these discrete spots, where these discrete spots identified as regions
of the high joint density of states are accommodated at around the scattering wave
vectors ${\bf q}_{i}$ obeying the {\it octet scattering model}
\cite{Hussey02,Balatsky06,Alloul09,Pan00,Hoffman02,McElroy03,Hanaguri07,Kohsaka08,Hanaguri09,Lee09,Vishik09,Schmidt11,Fujita14}
as shown in Fig. \ref{EFS-map}. This is why the octet scattering model can give a
consistent explanation of the STM/S experimental data
\cite{Hussey02,Balatsky06,Alloul09,Pan00,Hoffman02,McElroy03,Hanaguri07,Kohsaka08,Hanaguri09,Lee09,Vishik09,Schmidt11,Fujita14};
(iv) the LDOS pattern in momentum-space exhibits a particularly obvious band of the
scattering wave vectors in the diagonal direction forming a crisscrossing streaks
in the center of BZ, which arise from inter-tip scattering processes between states
on one Fermi arc. These diagonal streaks have a scattering wave vector length that is
set by $\bm{q}_{7}$. In particular, the QSI peaks with the scattering wave vectors
$\bm{q}_{2}$, $\bm{q}_{5}$, and $\bm{q}_{6}$ are broadened, while the QSI peaks
with the scattering wave vectors $\bm{q}_{1}$, $\bm{q}_{3}$, $\bm{q}_{4}$, and
$\bm{q}_{7}$ are most dominant. Moreover, the weights of the QSI peaks at the
scattering wave vectors $\bm{q}_{1}$ and $\bm{q}_{4}$ are smaller than the weights
of the QSI peaks at the scattering wave vectors $\bm{q}_{3}$ and $\bm{q}_{7}$,
indicating that the scattering processes with the sign-preserving scattering wave
vectors are suppressed;
(v) although the features of LDOS and of its Fourier-transformed form are quite
similar in the optimally doped and overdoped regimes, the behaviors of LDOS and of
its Fourier-transformed form in the underdoped regime are different from these in
the optimally doped and overdoped regimes, where (a) in the vicinity of the
impurity site, several isolated black islands and white regions appear, which are
corresponding to the high and low intensities of LDOS, respectively. In particular,
the intensities of the LDOS modulations both in real-space and momentum-space are
enhanced along the parallel directions; (b) the longer diagonal tails in LDOS
appeared in the optimally doped and overdoped regimes are suppressed in the
underdoped regime, and then disappear at the slightly underdoped region.
\begin{figure}[h!]
\centering
\includegraphics[scale=0.28]{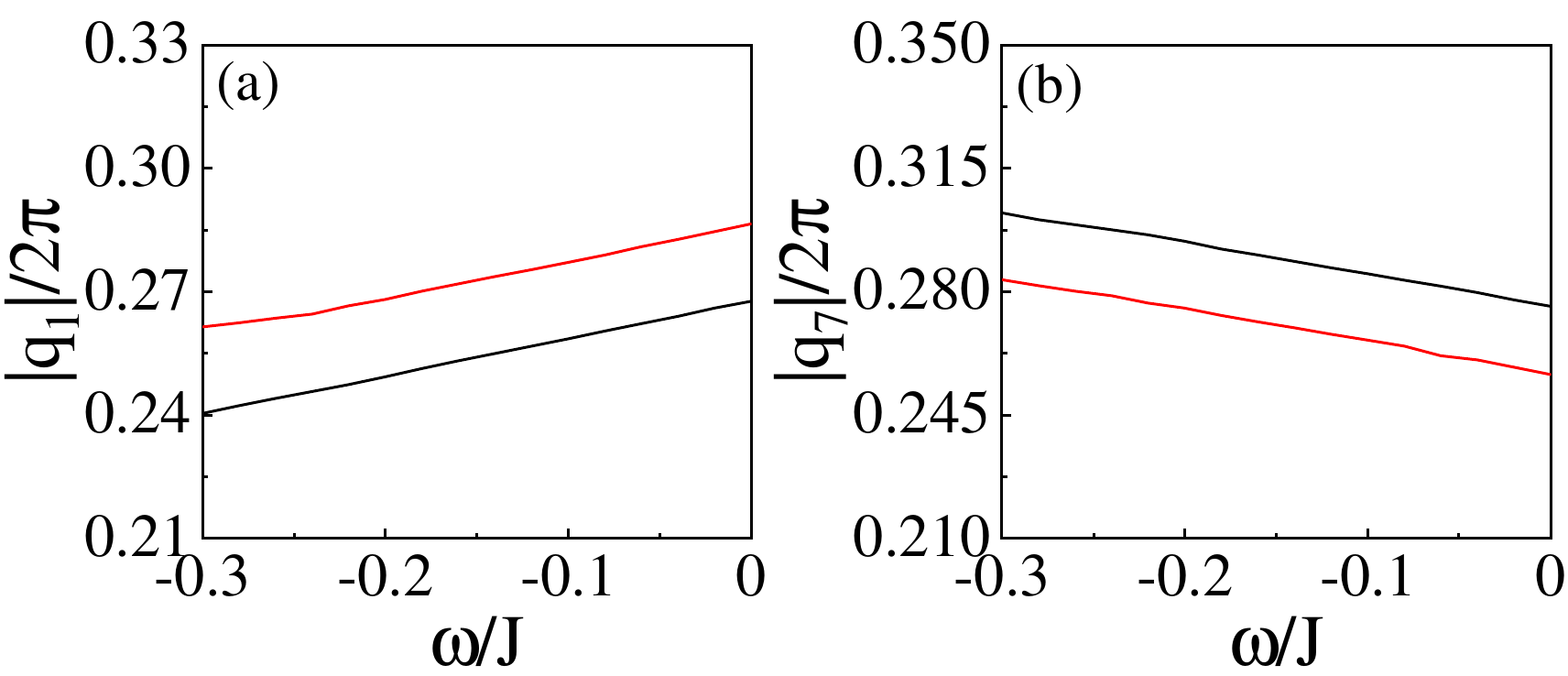}
\caption{(Color online) The scattering wave vectors (a) ${\bf q}_{1}$ and (b)
${\bf q}_{7}$ as a function of energy at $\delta=0.09$ (red-line) and $\delta=0.15$
(black-line) with $T=0.002J$ for the impurity-scattering strength $V_{s}=8J$ and
screening length $L=14a$.
\label{LDOS-MS-E}}
\end{figure}
On the
other hand, the Fourier-transformed LDOS in the diagonal direction displays a more
incoherent behavior; (c) in the region that is not far from the impurity site, the
intensity of the LDOS modulation along the diagonal direction is gradually increased
as the increase of doping, resulting in the enhancement of the intensity of the
Fourier-transformed LDOS along the diagonal direction;
(vi) in momentum-space, the regions displaying LDOS
modulation intensity change with energy, indicating that these modulations have
appreciable dispersion.
To characterize typical QSI peak dispersions more clearly, we plot the energy
dependence of the QSI peaks for the scattering wave vectors (a) ${\bf q}_{1}$ and
(b) ${\bf q}_{7}$ at $\delta=0.09$ (red-line) and $\delta=0.15$ (black-line) with
$T=0.002J$ for the impurity-scattering strength $V_{s}=8J$ and screening length
$L=14a$ in Fig. \ref{LDOS-MS-E}, where for a given doping concentration, the peak
at the scattering wave vector ${\bf q}_{1}$ evolves very differently with energy
than the peak at the scattering wave vector ${\bf q}_{7}$, and their dispersions
have opposite sign, i.e., the length of ${\bf q}_{1}$ becomes shorter, whereas the
length of ${\bf q}_{7}$ becomes longer, as energy $|\omega|$ is increased;
(vii) however, with the increase of the doping concentration, ${\bf q}_{1}$ decreases
systematically in length, while ${\bf q}_{7}$ moves to higher values, which are
expected since the distance between the parallel tips of the Fermi arcs moves closer
together, while the distance between the diagonal tips of the Fermi arcs increases
\cite{Zeng22,Gao19}. The results of the evolution of the QSI peaks with energy and
doping in Fig. \ref{LDOS-MS-E} therefore show that the typical QSI peak dispersions
are internally consistent within the octet scattering model shown in
Fig. \ref{EFS-map}. Our above results in Fig. \ref{LDOS-IPSI} and Fig. \ref{LDOS-MS-E}
are in qualitative agreement with the STM/S experimental observations
\cite{Hussey02,Balatsky06,Alloul09,Pan00,Hoffman02,McElroy03,Hanaguri07,Kohsaka08,Hanaguri09,Lee09,Vishik09,Schmidt11,Fujita14},
and are also in qualitative agreement with the corresponding ARPES experimental
results\cite{Chatterjee06,McElroy06,Chatterjee07,Restrepo23,He14}, where the doping
and energy dependence of the positions of the tips of the Fermi arcs (then the
scattering wave vectors ${\bf q}_{i}$) have been observed. In particular, the ARPES
experiments\cite{Comin16,Comin14} have indicated clearly that the magnitude of the
charge-order wave vector $Q_{\rm co}={\bf q}_{1}$ smoothly decreases with the increase
of doping, which is also in agreement with the present result shown in Fig. \ref{LDOS-MS-E}

In the above discussions, the crucial role of the strong electron correlation for
the LDOS modulation\cite{Comin16,Vishik18} is manifested by the electron self-energy
effect in the homogeneous electron propagator in Eq. (\ref{EGF-NR}). To clarify this
crucial self-energy effect on the LDOS modulation more clearly, we now discuss QSI
in the case of the absence of the electron self-energy, where the homogeneous
part of the SC-state electron propagator can be expressed as\cite{Feng15a},
\begin{equation}\label{RMGF-BCS}
\tilde{G}({\bf k},\omega)={Z_{\rm F}\over \omega^{2}-E_{\bf k}^{2}}\left (
\begin{array}{ll}\omega+\bar{\varepsilon}_{\bf k} &,-\bar{\Delta}_{\rm F}({\bf k}) \\
-\bar{\Delta}_{\rm F}({\bf k}) &,\omega-\bar{\varepsilon}_{\bf k} \end{array} \right),
\end{equation}
where
$E_{\bf k}=\sqrt{\bar{\varepsilon}_{\bf k}^2+\bar{\Delta}_{\rm F}^{2}({\bf k})}$,
the renormalized energy dispersion
$\bar{\varepsilon}_{\bf k}=Z_{\rm F}\varepsilon_{\bf k}$, and the renormalized SC
gap $\bar{\Delta}_{\rm F}({\bf k})=Z_{\rm F}\bar{\Delta}\gamma^{\rm (d)}_{\bf k}$,
with the d-wave factor $\gamma^{\rm (d)}_{{\bf k}}=(\cos k_{x}-\cos k_{y})/2$, while
the SC gap $\bar{\Delta}$ and the single-particle coherent weight $Z_{\rm F}$
have been given explicitly in Ref. \onlinecite{Feng15a}.

\begin{figure}[h!]
\centering
\includegraphics[scale=0.20]{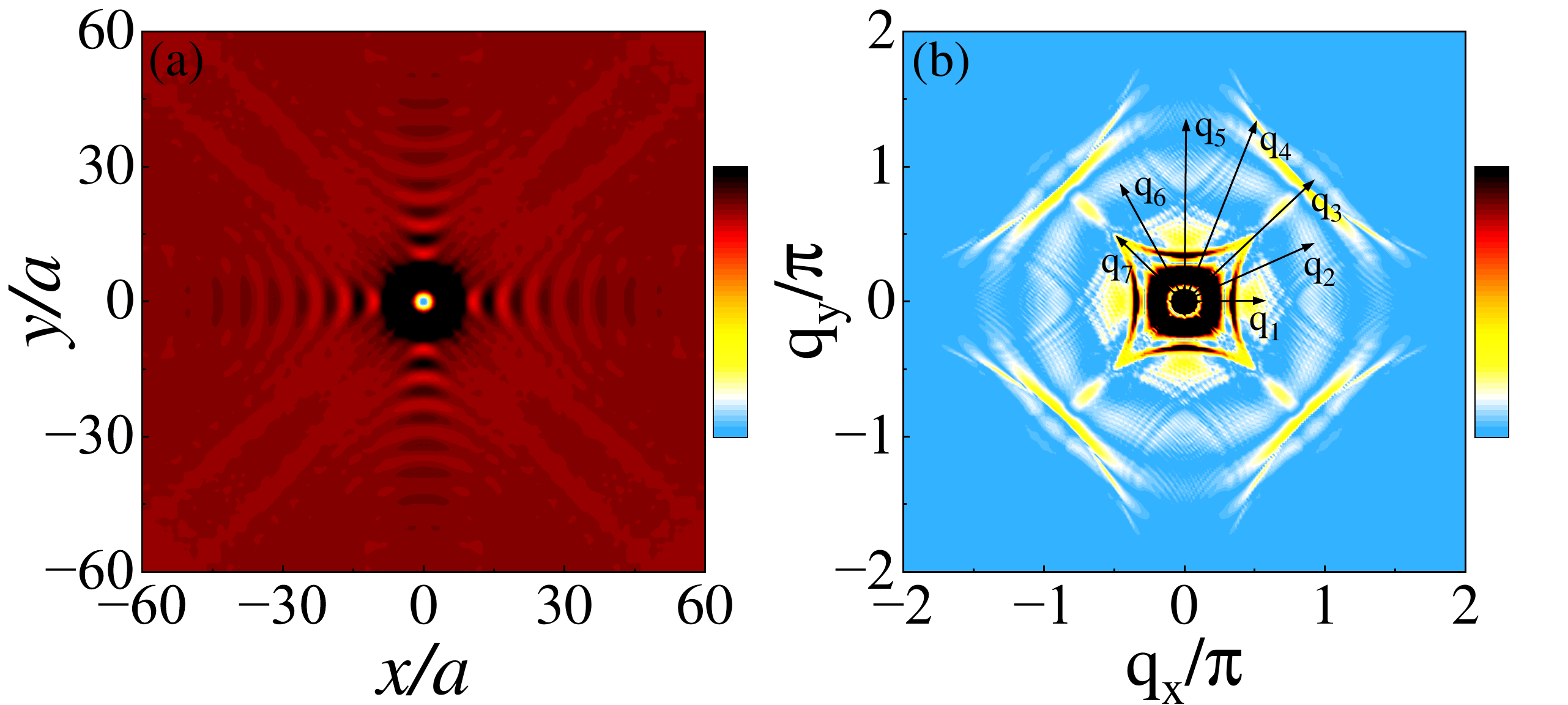}
\caption{(Color online) Maps of the local density of states in (a) real-space and
(b) momentum-space for an in-plane single impurity in the case of the absence of the
electron self-energy at $\omega=-0.16J$ and $\delta=0.15$ with $T=0.002J$ for
the impurity-scattering strength $V_{s}=8J$ and screening length $L=14a$.
\label{LDOS-BCS}}
\end{figure}

In Fig. \ref{LDOS-BCS}, we plot the maps of (a) the LDOS modulation and of (b) its
Fourier-transformed form in the case of the absence of the electron self-energy
for an in-plane single impurity at $\omega=-0.16J$ and $\delta=0.15$ with
$T=0.002J$ for the impurity-scattering strength $V_{s}=8J$ and screening length
$L=14a$. In comparison with the corresponding results shown in Fig. \ref{LDOS-IPSI}b
and Fig. \ref{LDOS-IPSI}e, it is easy to find that there are some substantial
differences between the cases of the presence and absence of the electron
self-energy. In particular, these substantial differences can be summarized
as: (i) the LDOS modulation is particularly pronounced along the parallel direction,
where the separation between the bright streaks remain unchanged even for the
distances of these streaks that are far from the scatterer. Moreover, the weights
of the long streaks along the diagonal direction are weaker than these along
the parallel direction;
(ii) the weights of the QSI peaks in momentum-space are more
dispersive, and more specially, the weight of the QSI peak with the wave vector
${\bf q}_1$ along the parallel direction is much larger than that of the QSI peak
with the wave vector ${\bf q}_7$ along the diagonal direction, which are confronted
with the real-space LDOS pattern, however, they are inconsistent with the
corresponding results shown in Fig. \ref{LDOS-IPSI}b and Fig. \ref{LDOS-IPSI}e,
where the results are in agreement with the experiment data
\cite{Hussey02,Balatsky06,Alloul09,Pan00,Hoffman02,McElroy03,Hanaguri07,Kohsaka08,Hanaguri09,Lee09,Vishik09,Schmidt11,Fujita14};
(iii) although the QSI peaks
with the characteristic wave vectors ${\bf q}_{i}$ in Fig. \ref{LDOS-BCS}b are still
discernible, they are too dispersive to be as well-defined as these shown in
Fig. \ref{LDOS-IPSI}e. These results in Fig. \ref{LDOS-BCS} are qualitatively
consistent with these in the similar discussions
\cite{Capriotti03,Wulindan10,Nowadnick12,Torre16,Torre16NP,Wang03,Zhang03,Zhu04a,Wangshuhua15,ZhaoMM23,Kreisel15,Choubey17,Sulangi17,Sulangi18-a,Sulangi18-b}.
Incorporating both the results in Fig. \ref{LDOS-BCS} and
Fig. \ref{LDOS-IPSI}, it is thus shown that the essential features of the QSI peaks
with the wave vectors ${\bf q}_{i}$ obtained due to the inclusion of the strong
electron correlation effect in the homogeneous electron propagator are consistent
with the experimental observations, confirming that the inhomogeneity is intrinsic
to the electronic structure\cite{Comin16,Vishik18,Pan01,Vershinin04}.
\begin{figure*}[t!]
\centering
\includegraphics[scale=0.205]{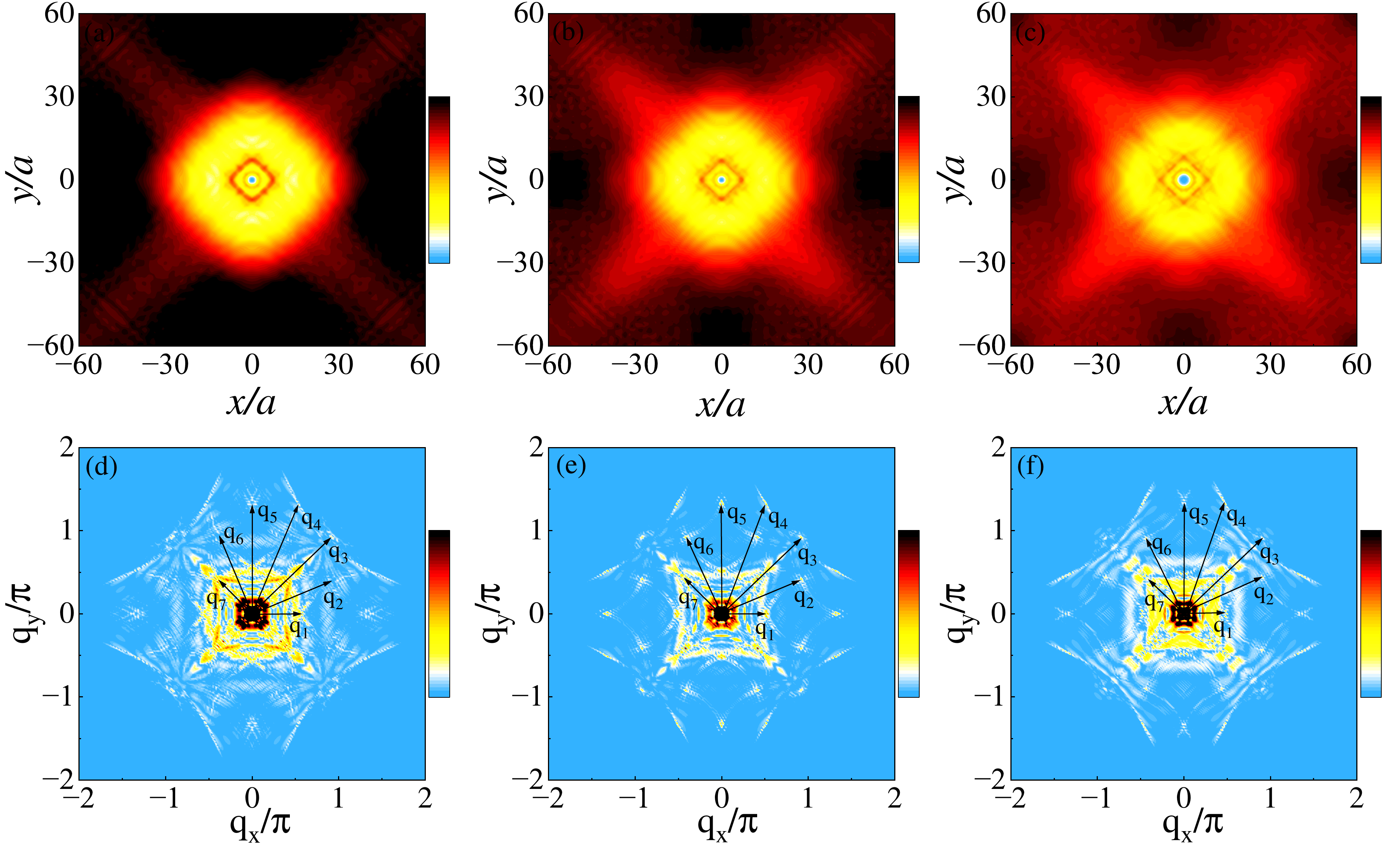}
\caption{(Color online) Upper panel: the maps of the local density of states in
real-space for an out-of-plane single impurity at (a) $\delta=0.09$, (b)
$\delta=0.15$, and (c) $\delta=0.21$ in $\omega=-0.16J$ and $T=0.002J$ with the
impurity-scattering strength $V_{s}=5J$ and screening length $L=5a$.
Lower panel: the corresponding maps of the local density of states in
momentum-space at (d) $\delta=0.09$, (e) $\delta=0.15$, and (f) $\delta=0.21$
Fourier-transformed from the maps of the real-space local density of states in
(a), (b), and (c), respectively.
\label{LDOS-OPSI}}
\end{figure*}

\subsection{Out-of-plane single impurity}

Now we turn to study the characteristic features of QSI arising from an
out-of-plane single impurity. In the out-of-plane impurity system
\cite{Eisaki04,Fujita05,McElroy05,Kondo07,Hashimoto08}, the dopants are charged
impurities which act as sources for poorly screening Coulomb potentials, which in
turn affect the essential physics on the CuO$_{2}$ planes\cite{Pan01,Sulangi17}. In
this case, the scattering is relatively weak, and then the out-of-plane single
impurity-scattering potential in real-space can be modeled as\cite{Zhu04,Nunner06},
\begin{equation}\label{VPOT-2-1}
V(r) =V_{s}e^{-r/L},
\end{equation}
with $V_{s}$ and $L$ being the impurity-scattering strength and screening length,
respectively, while its Fourier transformed form can be derived directly as,
\begin{equation}\label{VPOT-2}
V_{\bm{k}\bm{k}'}={2\pi V_{s}/L\over [(\bm{k}-\bm{k}')^{2}+1/L^{2}]^{3/2}}.
\end{equation}

In the upper panel of Fig. \ref{LDOS-OPSI}, we plot the LDOS maps in real-space
for an out-of-plane single impurity at (a) $\delta=0.09$, (b) $\delta=0.15$, and
(c) $\delta=0.21$ in $\omega=-0.16J$ and $T=0.002J$ with the impurity-scattering
strength $V_{s}=5J$ and screening length $L=5a$, while the corresponding LDOS maps
in momentum-space at (d) $\delta=0.09$, (e) $\delta=0.15$, and (f) $\delta=0.21$
are plotted in the lower panel of Fig. \ref{LDOS-OPSI}, which are obtained from
the real-space LDOS maps in (a), (b), and (c), respectively, in terms of the
Fourier-transform.
In comparison with the corresponding results obtained for an in-plane single
impurity shown in Fig. \ref{LDOS-IPSI}, it thus shows that the main feature of the
LDOS modulation for an out-of-plane single impurity is similar to that for an
in-plane single impurity. In particular, as a natural consequence of the
quasiparticle scattering geometry that is directly governed by the shape of the
constant energy contour and the related octet scattering model shown in
Fig. \ref{EFS-map}, (i) the scattering wave vectors $\bm{q}_{i}$ obey the octet
scattering model, in spite of various forms of the impurity scattering potentials;
(ii) the features of the doping dependence of QSI are similar to these for an
in-plane single impurity as shown in Fig. \ref{LDOS-IPSI};
(iii) the behaviours of the doping dependence of the QSI peak dispersions for an
out-of-plane single impurity are almost the same as these for an in-plane single
impurity as shown in Fig. \ref{LDOS-MS-E}, which also coincide with the evolution
of the constant energy contour and the related octet scattering model in
Fig. \ref{EFS-map} with energy and doping concentration. However, there are some
subtle differences between the LDOS modulations for an out-of-plane single impurity
and an in-plane single impurity: (i) the complicated real-space structure of the
LDOS modulation leads to that the momentum-space structure of the LDOS modulation is
more pronounced, where except for the pronounced $\bm{q}_{1}$, $\bm{q}_{3}$,
$\bm{q}_{4}$, and $\bm{q}_{7}$ peaks, the $\bm{q}_{2}$, $\bm{q}_{5}$, and
$\bm{q}_{6}$ peaks that are broadened for an in-plane single impurity are clearly
discernible for an out-of-plane single impurity, and then all the QSI peaks with the
scattering wave vectors $\bm{q}_{i}$ are the dominant characteristic of the LDOS
modulation in momentum-space; (ii) the shape of the LDOS modulation pattern for an
out-of-plane single impurity deviates dramatically from the square-like shape of
the LDOS modulation pattern for an in-plane single impurity shown in
Fig. \ref{LDOS-IPSI}; As we have mentioned in Eqs. (\ref{Rhoq-R}) and (\ref{Rhoq}),
the present results in Fig. \ref{LDOS-IPSI} and Fig. \ref{LDOS-OPSI} therefore show
that various forms of the impurity-scattering potentials can induce some subtle
differences for the LDOS modulation.
In particular, the QSI peaks in momentum-space from the out-of-plane single impurity
scattering are more pronounced than these from the in-plane single impurity
scattering, indicating that QSI in cuprate superconductors may be induced mainly by
the out-of-plane single impurity scattering.
However, the in-plane and
out-of-plane impurities may coexist in cuprate superconductors, which leads to
that the LDOS modulation varies from samples to samples, although the main features
of the momentum-space LDOS modulation are universal for all cuprate superconductors.
In this case, the STM/S experimental data may be well explained by making use of the
specific combination of the in-plane and out-of-plane impurity scattering potentials.

To further explore the similarities and differences between the LDOS modulation for
an out-of-plane single impurity and an in-plane single impurity, we introduce the
{\it impurity weight}\cite{Sulangi17}, which can be expressed in the present case
as,
\begin{equation}\label{imp-weight-1}
W_{0}(V_{s},L,\omega)={\sum \limits_{\bm{q}}|
\rho(\bm{q},V_{s},L,\omega)|-|\rho(\bm{q}_{0},V_{s},L,\omega)|\over
\sum \limits_{\bm{q}}|\rho(\bm{q},V_{s},L,\omega)|},
\end{equation}
where $\bm{q}_{0}=[0,0]$ is the point of the brightest spot located at the center
of BZ, while the summation of momentum ${\bf q}$ is restricted within
$[-2\pi\leq q_{x}\leq 2\pi]$ and $[-2\pi\leq q_{y}\leq 2\pi]$.
This impurity weight is a ratio of the integrated LDOS spectrum in momentum-space
without the $\bm{q}_{0}$ contribution to the total integrated LDOS spectrum, namely,
$\rho(\bm{q}_{0},V_{s},L,\omega)$ is removed from the numerator since its
contribution originates from the Fourier-transform of the LDOS for a spatially
homogeneous cuprate superconductor with the d-wave symmetry. In this case, the
numerator of Eq. (\ref{imp-weight-1}) therefore depicts only the contributions of
the inhomogeneities to the LDOS spectrum.
However, as we have mentioned in Sec. \ref{Formalism}, the calculation for the
normal and anomalous self-energies and the related LDOS is performed numerically on
a $120\times 120$ lattice in momentum space, with the infinitesimal
$i0_{+}\rightarrow i\Gamma$ replaced by a small damping $\Gamma=0.05J$, which leads
to that the weight of the central peak in Fig. \ref{LDOS-IPSI}b and
Fig. \ref{LDOS-OPSI}b at the wave vector $\bm{q}_{0}=[0,0]$ spreads on the small
area $\{{\bf Q}\}$ around the ${\bf q}_{0}$ point. In particular, the summation of
these spread weights around this small area $\{{\bf Q}\}$ is less affected by the
calculation for a finite lattice. In this case, a more appropriate {\it impurity
weight} can be defined as,
\begin{equation}\label{imp-weight}
W(V_{s},L,\omega)=1-{\sum_{{\bf q}\in\{{\bf Q}\}}|\rho(\bm{q},V_{s},L,\omega)|
\over\sum\limits_{\bm{q}}|\rho(\bm{q},V_{s},L,\omega)|},
\end{equation}
where the summation ${\bf q}\in \{{\bf Q}\}$ is restricted to the small area
$\{{\bf Q}\}$ around the ${\bf q}_{0}$ point. To see this impurity-scattering
strength and screening length dependence of the impurity weight more clearly, we
first plot the impurity weight $W(V_{s},L,\omega)$ as a function of the
impurity-scattering strength $V_{s}$ in $\omega=-0.16J$ at $\delta=0.15$ with
$T=0.002J$ and the screening length $L=14a$ for an in-plane single impurity (black
squares) and the screening length $L=5a$ for an out-of-plane single impurity (red
dots) in Fig. \ref{weight-V}, where at a given impurity-scattering strength, the
impurity weight for an in-plane impurity is always larger than that for an
out-of-plane impurity. In particular, the impurity-scattering strength dependence
of the impurity weight can be separated into three typical regions:
\begin{figure}[h!]
\centering
\includegraphics[scale=0.235]{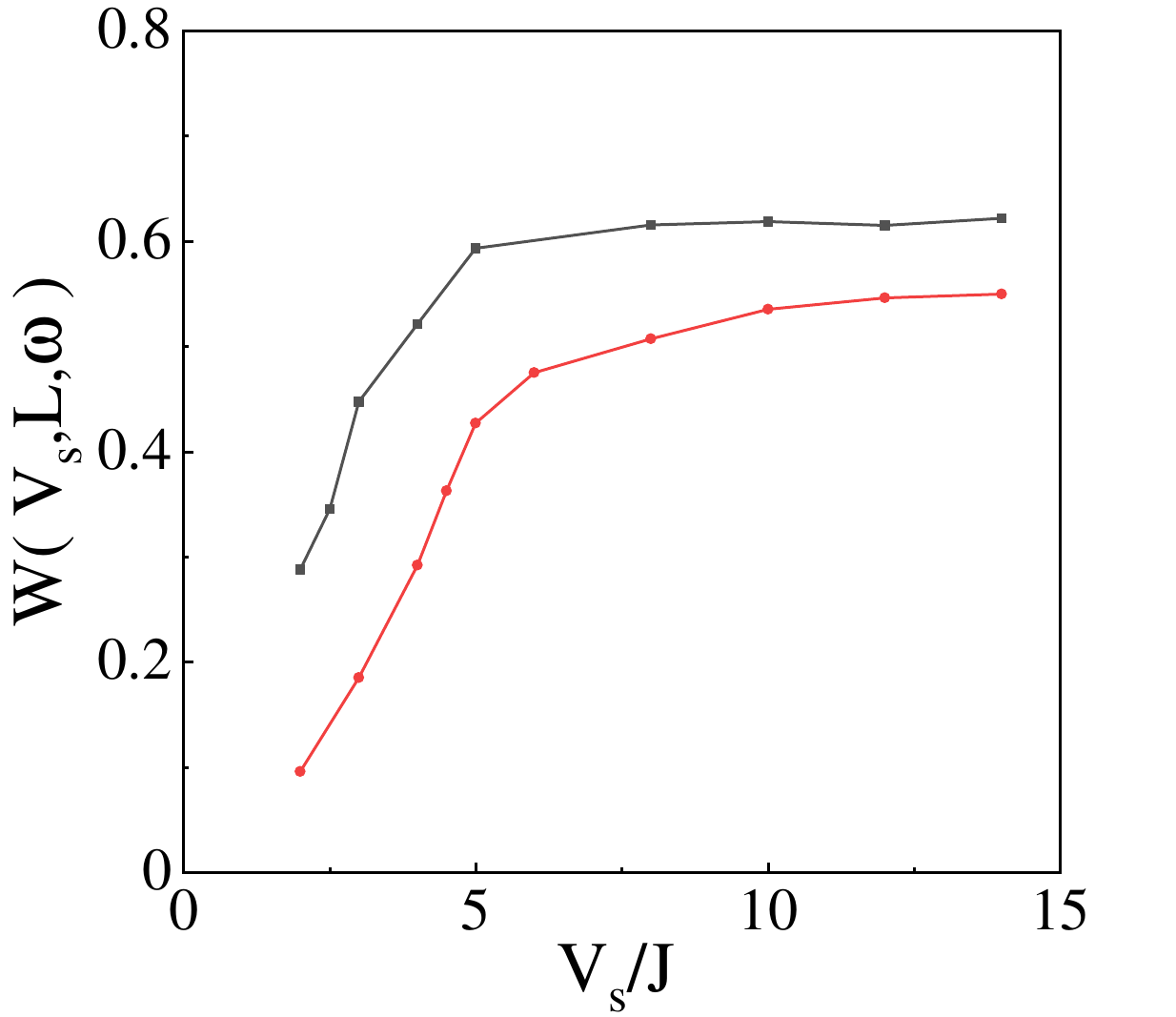}
\caption{(Color online) The impurity weight as a function of the impurity-scattering
strength $V_{s}$ in $\omega=-0.16J$ at $\delta=0.15$ with $T=0.002J$ and the
screening length $L=14a$ for an in-plane single impurity (black squares) and the
screening length $L=5a$ for an out-of-plane single impurity (red dots).
The black- and red-lines are guide for the eyes.\label{weight-V}}
\end{figure}
(i) in the
region of the weak impurity-scattering strength, which is corresponding to the
case of $V_{s} < 5J$, where the impurity weight depends almost linearly on the
impurity-scattering strength, i.e., it rises almost linearly as the
impurity-scattering strength is increased; (ii) in the crossover region, which is
corresponding to the case of $5J < V_{s} < 8J$, where the impurity weight increases
slowly with the increase of the impurity-scattering strength; (iii) however, in the
region of the strong impurity-scattering strength, which is corresponding to the
case of $V_{s} > 8J$, where the impurity weight tends to saturate to a fixed value
$\sim 0.6$ for an in-plane impurity and a fixed value $\sim 0.5$ for an out-of-plane
impurity. These results in Fig. \ref{weight-V} are well consistent with the results
obtained based on the real-space Green's function approach\cite{Sulangi17}, and are
helpful to identify the nature of the in-plane or out-of-plane scatterer.

\begin{figure}[h!]
\centering
\includegraphics[scale=0.235]{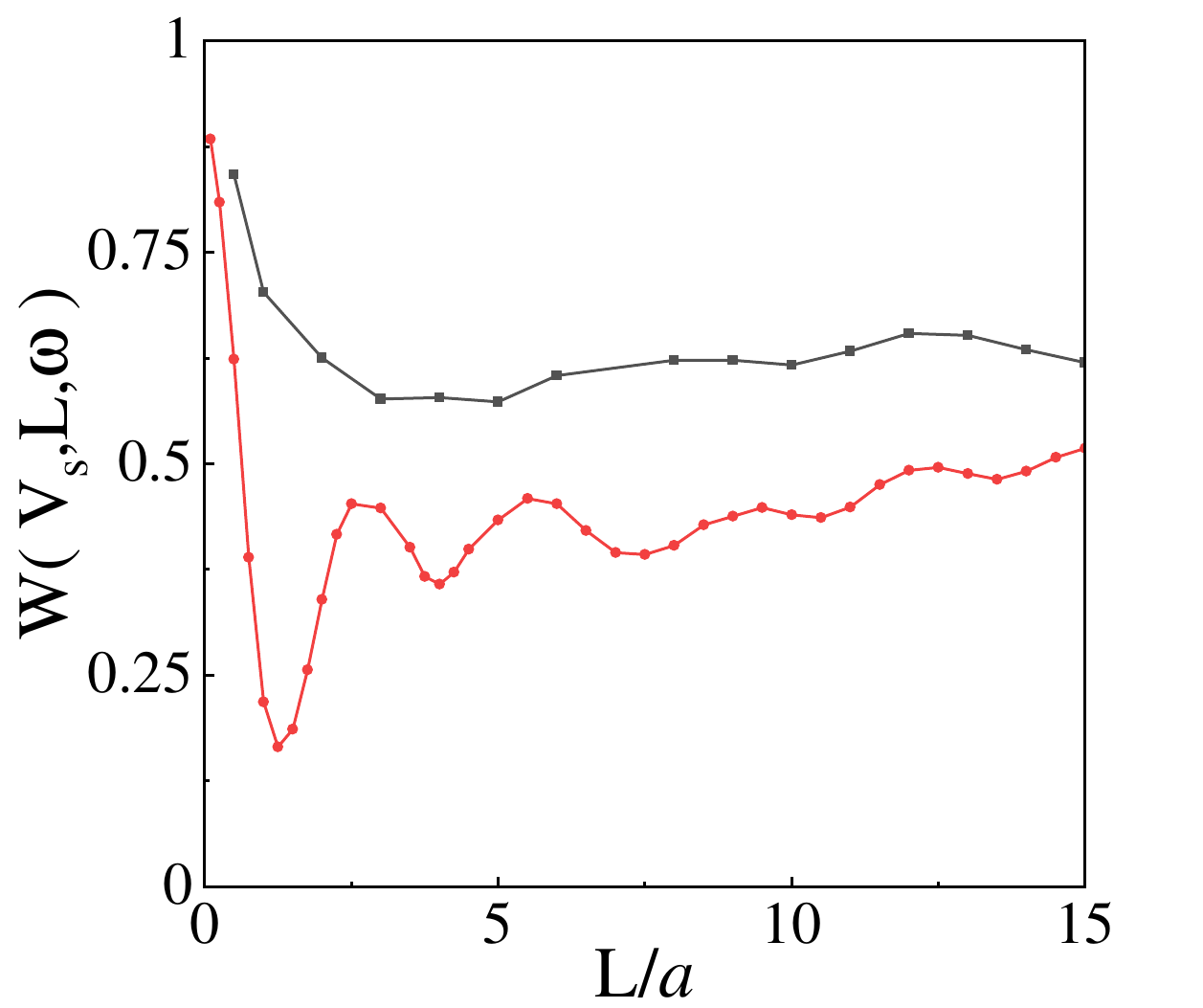}
\caption{(Color online) The impurity weight as a function of screening length $L$
for $\omega=-0.16J$ at $\delta=0.15$ with $T=0.002J$ and the impurity-scattering
strength $V_{s}=8J$ for an in-plane single impurity (black squares) and the
impurity-scattering strength $V_{s}=5J$ for an out-of-plane impurity (red dots).
The black- and red-lines are guide for the eyes.\label{weight-L}}
\end{figure}

From the LDOS modulation in Eqs. (\ref{Rhoq-R}) and (\ref{Rhoq}), it thus shows that
except for the homogeneous electron propagator, the structure of the LDOS modulation
is partially determined by the impurity-scattering potential in the $T$-matrix. In
other words, the distribution of the weight in the momentum-space LDOS modulation is
set partially by the characteristic wave vectors of the scattering potential, which
leads to that the structure of the momentum-space LDOS modulation is sensitive to
the length scale (then the screening length) associated with the impurity-scattering
potential. To see this point more clearly, we plot the impurity weight as a function
of screening length $L$ for $\omega=-0.16J$ at $\delta=0.15$ with $T=0.002J$ and the
impurity-scattering strength $V_{s}=8J$ for an in-plane single impurity (black
squares) and impurity-scattering strength $V_{s}=5J$ for an out-of-plane impurity
(red dots) in Fig. \ref{weight-L},
where for the case of the short screening length
($L < 2a$ for an in-plane single impurity and $L < 1a$ for an out-of-plane impurity),
the impurity weight decreases dramatically as the growth of the screening length,
and this trend is quite consistent with the behavior of the impurity-scattering
potentials in Eqs. (\ref{VPOT-1}) and (\ref{VPOT-2}) in the vicinity of
$\bm{q}=[0,0]$ as the screening length is increased. However, for the case beyond
the short screening length, the behaviour of the impurity weight for an in-plane
single impurity is quite different from that for an out-of-plane single impurity,
where the impurity weight for an in-plane single impurity saturates to a fixed
value, while it for an out-of-plane single impurity oscillates with the increase
of screening length, and saturates to a fixed value for the long screening length
$L > 15a$. It has been shown that for a long screening length $L$, the
momentum-space LDOS modulation spectrum is associated with small-momentum
scattering processes, whereas for a short screening length $L$, the more spectral
weight of the momentum-space LDOS modulation spectrum is associated with
large-momentum processes\cite{Sulangi17}. Our results in Fig. \ref{weight-L}
therefore indicate that the momentum-space LDOS modulation spectrum in
Fig. \ref{LDOS-IPSI}b for an in-plane single impurity with the screening length
$L=14a$ is mainly associated with small-momentum scattering processes, while the
momentum-space LDOS modulation spectrum in Fig. \ref{LDOS-OPSI}b for an out-of-plane
single impurity with the screening length $L=5a$ hosts the contributions from both
small- and large-momentum processes. This is why there are some substantial
differences between the momentum-space LDOS modulations for an out-of-plane single
impurity in Fig. \ref{LDOS-OPSI}b and an in-plane single impurity in
Fig. \ref{LDOS-IPSI}b. In particular, the results in Fig. \ref{weight-L} also show
that only in the case of very short screening length, both small- and
large-momentum processes figure prominently in the momentum-space LDOS modulation
spectrum, and then the features of the momentum-space LDOS modulation spectrum for
an out-of-plane single impurity resembles that for an in-plane single impurity.

\begin{figure*}[t!]
\centering
\includegraphics[scale=0.20]{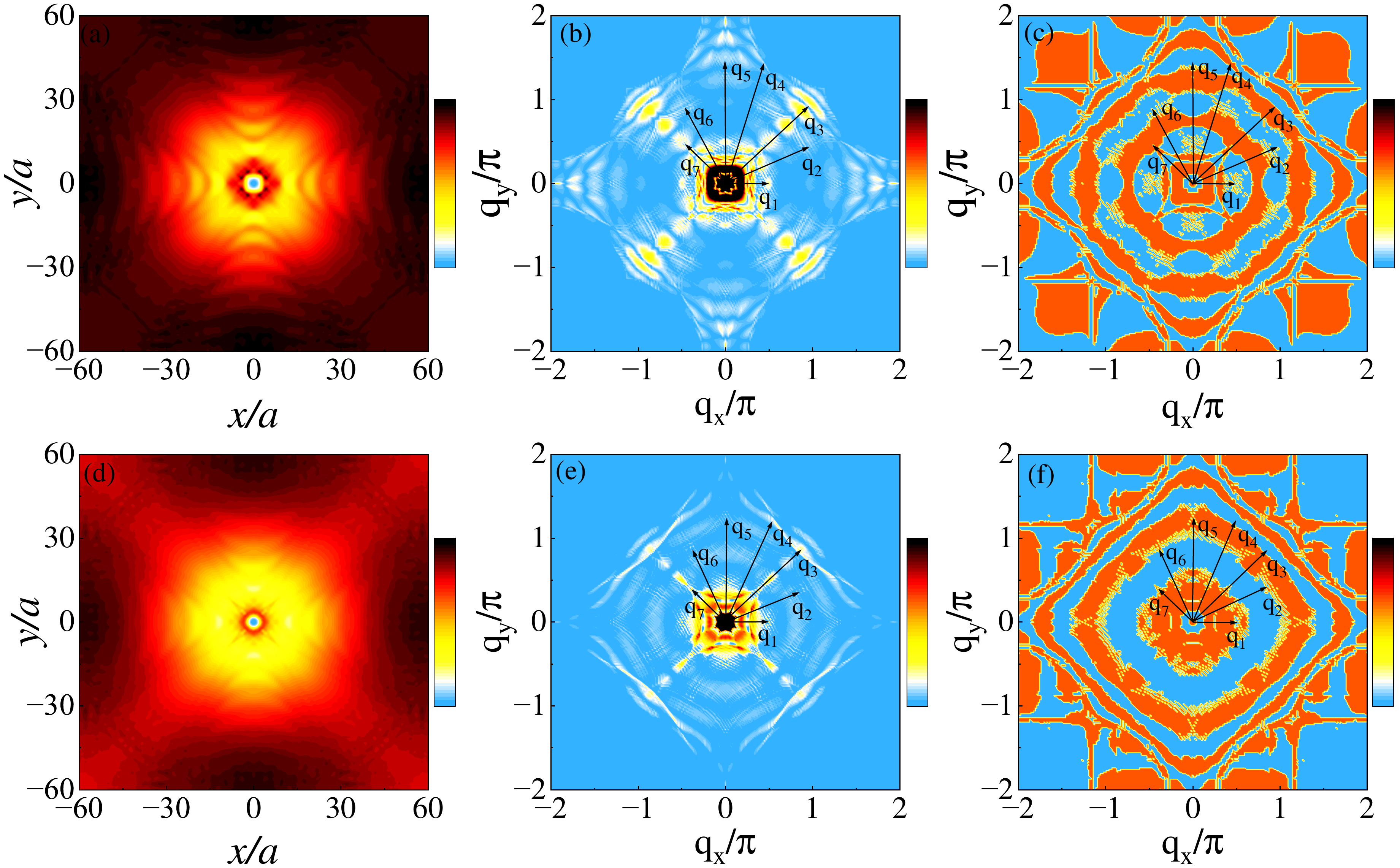}
\caption{(Color online) Upper panel: the maps of (a) the filtered local density of
states in real-space, (b) the amplitude of the filtered local density of states
in momentum-space, and (c) the phase of the filtered local density of states for an
in-plane single impurity in $\omega=-0.16J$ at $\delta=0.15$ with $T=0.002J$ for
the impurity-scattering strength $V_{s}=8J$ and the screening length $L=14a$.
Lower panel: the corresponding maps of (d) the local density of states in real-space,
(e) the amplitude of the local density of states in momentum-space, and (f) the
phase of the local density of states in the case of the absence of the filter effect
for an in-plane single impurity. \label{LDOS-bare-filtered-in}}
\end{figure*}

\subsection{Filter effect on the electronic inhomogeneity}
\label{Results-FilterLDOS}

The {\it filter effect} is the effect, where it is assumed that the tunneling
process from the tip to the CuO$_{2}$ plane involves the atomic wave functions in
the neighboring unit cells\cite{Torre16,Torre16NP,JXZhu2000,Martin2002}, which is
due to the mismatch between the s-wave orbital from the STM/S tip and the d-wave
copper orbitals in the CuO$_{2}$ plane. This filter effect on the electronic
inhomogeneity has been studied widely, however, the conclusions are diverse. On the
one hand, the obtained results of the LDOS modulation based on the combination of
the Bogoliubov-de Gennes equaiton and Wannier function from a first-principle
calculation show the importance of the filter effect on the electronic inhomogeneity
\cite{Kreisel15}. In particular, it has been shown that the tunneling matrix element
has actually a d-wave symmetry, because the most dominant tunneling processes are
arisen from the d-wave copper orbital of the NN sites. In this case, the filter
effect on the LDOS modulation comes mainly from the contributions from the four NN
sites\cite{Martin2002}. On the other hnad, the simplest form of the filter function
$f({\bf r},{\bf r}')$ has been employed to discuss the filter effect on the
electronic inhomogeneity\cite{Choubey17}, and the results show that the
{\it filter effect} only induces a shift of the spectral weight from one part of
momentum-space to another, while the locations of the QSI peaks predicted by the
octet model remain unchanged, indicating that the filter effect has a small
influence on the LDOS modulation.

In this subsection, we study the filter effect on the LDOS modulation in terms of
the impurity-dressed electron propagator $\tilde{G}_{I}({\bf r},{\bf r}',\omega)$
in Eq. (\ref{Green-Fun-k-1}) and the filter function\cite{Sulangi17}
$f({\bf r},{\bf r}')$. Following the discussion in Ref. \onlinecite{Sulangi17}, the
filtered LDOS $\rho_{\rm f}({\bf r},\omega)$ can be expressed as,
\begin{equation}
\rho_{\rm f}({\bf r},\omega)=-{2\over\pi}{\rm Im}\sum_{{\bf r}_{1},{\bf r}_{2}}
f({\bf r}-{\bf r}_{1},{\bf r}-{\bf r}_{2})
\tilde{G}_{I}({\bf r}_{1},{\bf r}_{2},\omega)_{11},
\end{equation}
with the filter function,
\begin{eqnarray}\label{Filter-Fun}
f({\bf r},{\bf r}')&=&(\delta_{{\bf r},\hat{x}}+\delta_{{\bf r},-\hat{x}}
-\delta_{{\bf r},\hat{y}}-\delta_{{\bf r},-\hat{y}}) \nonumber\\
&\times&(\delta_{{\bf r}',\hat{x}}+\delta_{{\bf r}',-\hat{x}}
-\delta_{{\bf r}',\hat{y}}-\delta_{{\bf r}',-\hat{y}}),
\end{eqnarray}
where $\hat{x}$ and $\hat{y}$ are unit vectors in the $x$ and $y$ directions,
respectively. From the impurity-dressed electron propagator
$\tilde{G}_{I}({\bf r},{\bf r}',\omega)$ in Eq. \ref{Green-Fun-k-1}, the filtered
LDOS in real-space can be derived as,
\begin{eqnarray}\label{Filter-LDOS}
&&\rho_{f}({\bf r},\omega)=\rho_{f}^{(0)}(\omega)+\delta\rho_{f}({\bf r},\omega)
\nonumber\\
&&=-{32\over N\pi}{\rm Im}\big\{ \sum_{{\bf k}}\tilde{G}({\bf k},\omega)
{\gamma^{\rm (d)}_{{\bf k}}}^{2} \nonumber\\
&&+ \sum_{{\bf k},{\bf k}'}\tilde{G}({\bf k},\omega)
\tilde{T}({\bf k},{\bf k}',\omega)\tilde{G}({\bf k}',\omega)e^{i({\bf k}-{\bf k}')
\cdot{\bf r}}\gamma^{\rm (d)}_{{\bf k}}\gamma^{\rm (d)}_{{\bf k}'}\big\}_{11}.
\nonumber\\
\end{eqnarray}
With the help of the Fourier transform form of the filter function,
\begin{equation}\label{d-wave-factor}
f({\bf k},{\bf k}')=(8/N)\gamma^{\rm (d)}_{\bf k}\gamma^{\rm (d)}_{{\bf k}'},
\end{equation}
and removing the contribution from zero momentum, the Fourier transformed form of
the filtered LDOS now can be obtained explicitly as,
\begin{eqnarray}\label{Filter-rhofqresult}
\delta\rho_{f}({\bf q},\omega)&=&{\rm Re}[\delta\rho_{f}({\bf q},\omega)]
+i{\rm Im}[\delta\rho_{f}({\bf q},\omega)],
\end{eqnarray}
where the real part ${\rm Re}[\delta\rho_{f}({\bf q},\omega)]$ and imaginary part
${\rm Im}[\delta\rho_{f}({\bf q},\omega)]$ of the filtered LDOS in momentum-space
$\delta\rho_{f}({\bf q},\omega)$ are given by,
\begin{subequations}
\begin{eqnarray}
{\rm Re}[\delta\rho_{f}({\bf q},\omega)]&=&-{2\over\pi}{\rm Im}
[\tilde{A}_{f}({\bf q},\omega)+\tilde{A}_{f}(-{\bf q},\omega)]_{11},~~~~~
\label{Filter-rhofqresult-RI}\\
{\rm Im}[\delta\rho_{f}({\bf q},\omega)]&=&-{2\over\pi}{\rm Re}
[\tilde{A}_{f}({\bf q},\omega)-\tilde{A}_{f}(-{\bf q},\omega)]_{11}, ~~~~~
\label{Filter-rhofqresult-RI}
\end{eqnarray}
\end{subequations}
respectively, with the function,
\begin{equation}
\tilde{A}_{f}({\bf q},\omega)=4\sum_{{\bf k}}\gamma^{\rm (d)}_{\bf k}
\gamma^{\rm (d)}_{{\bf k}+{\bf q}}\tilde{G}({\bf k},\omega)
\tilde{T}({\bf k},{\bf k}+{\bf q},\omega)\tilde{G}({\bf k}+{\bf q},\omega).
\end{equation}
In this case, the amplitude of the filtered LDOS in momentum-space and the related
phase (then angle $\phi$ of the phase factor) can be obtained explicitly as,
\begin{subequations}\label{Amplitude-phase}
\begin{eqnarray}
|\delta\rho_{f}({\bf q},\omega)|&=&
\sqrt{({\rm Re}[\delta\rho_{f}({\bf q},\omega)])^{2}
+({\rm Im}[\delta\rho_{f}({\bf q},\omega)])^{2}},~~~~~~~~\label{Amplitude-LDOS} \\
\phi &=&\arctan\left ({{\rm Im}[\delta\rho_{f}({\bf q},\omega)]\over
{\rm Re}[\delta\rho_{f}({\bf q},\omega)]}\right )\nonumber\\
&+&\pi\theta(-{\rm Re}[\delta\rho_{f}({\bf q},\omega)]), ~~~~~\label{Phase-LDOS}
\end{eqnarray}
\end{subequations}
respectively, with the step function $\theta(\varepsilon)$. The above results in
Eqs. (\ref{Filter-LDOS}) and (\ref{Filter-rhofqresult}) show
clearly that in the case of the presence of the {\it filter effect}, although the
positions of the QSI peaks are unchanged, the partial spectral weight is moved
from one part to another\cite{Torre16,Torre16NP,Sulangi17}, where the amount of
the shifted spectral weight is mainly dependent on the d-wave factor
$\gamma^{\rm (d)}_{\bm{k}}$. To
see these features more clearly, we plot the maps of (a) the filtered LDOS in
real-space, (b) the amplitude of the filtered LDOS in momentum-space, and (c) the
phase of the filtered LDOS for an in-plane single impurity in $\omega=-0.16J$ at
$\delta=0.15$ with $T=0.002J$ for the impurity-scattering strength $V_{s}=8J$ and
the screening length $L=14a$ in the upper panel of Fig. \ref{LDOS-bare-filtered-in}.
For a better comparison, the corresponding maps of (d) LDOS in real-space, (e) the
amplitude of LDOS in momentum-space, and (f) the phase of LDOS in the case of the
absence of the filter effect for an in-plane single impurity are also plotted in the
lower panel of Fig. \ref{LDOS-bare-filtered-in}. It thus shows that (i) the role of
the {\it filter effect} actually leads to the filtered LDOS in real-space more fuzzy,
and the disappearance of some subtle details along the diagonal direction of the
filtered LDOS map. In particular, the d-wave factor in the filter function
(\ref{d-wave-factor}) induces the streaks in the filtered LDOS. Moreover, the
intensity of the filtered LDOS along the parallel direction are significantly
enhanced. However, in contrast to the case with the local on-site
impurity potential\cite{Torre16,Torre16NP,Kreisel15,Sulangi17}, the d-wave
factor in the filter function (\ref{d-wave-factor}) does not transform the local
minimum of the LDOS intensity on the impurity site into a local maximum for a
realistic impurity potential with finite screening lengths; (ii) although the QSI
peaks are still discernible in the filtered LDOS in momentum-space, these peaks are
broadened and are characterized with a dispersive pattern in momentum-space,
especially for these along the diagonal direction.
\begin{figure*}[t!]
\centering
\includegraphics[scale=0.20]{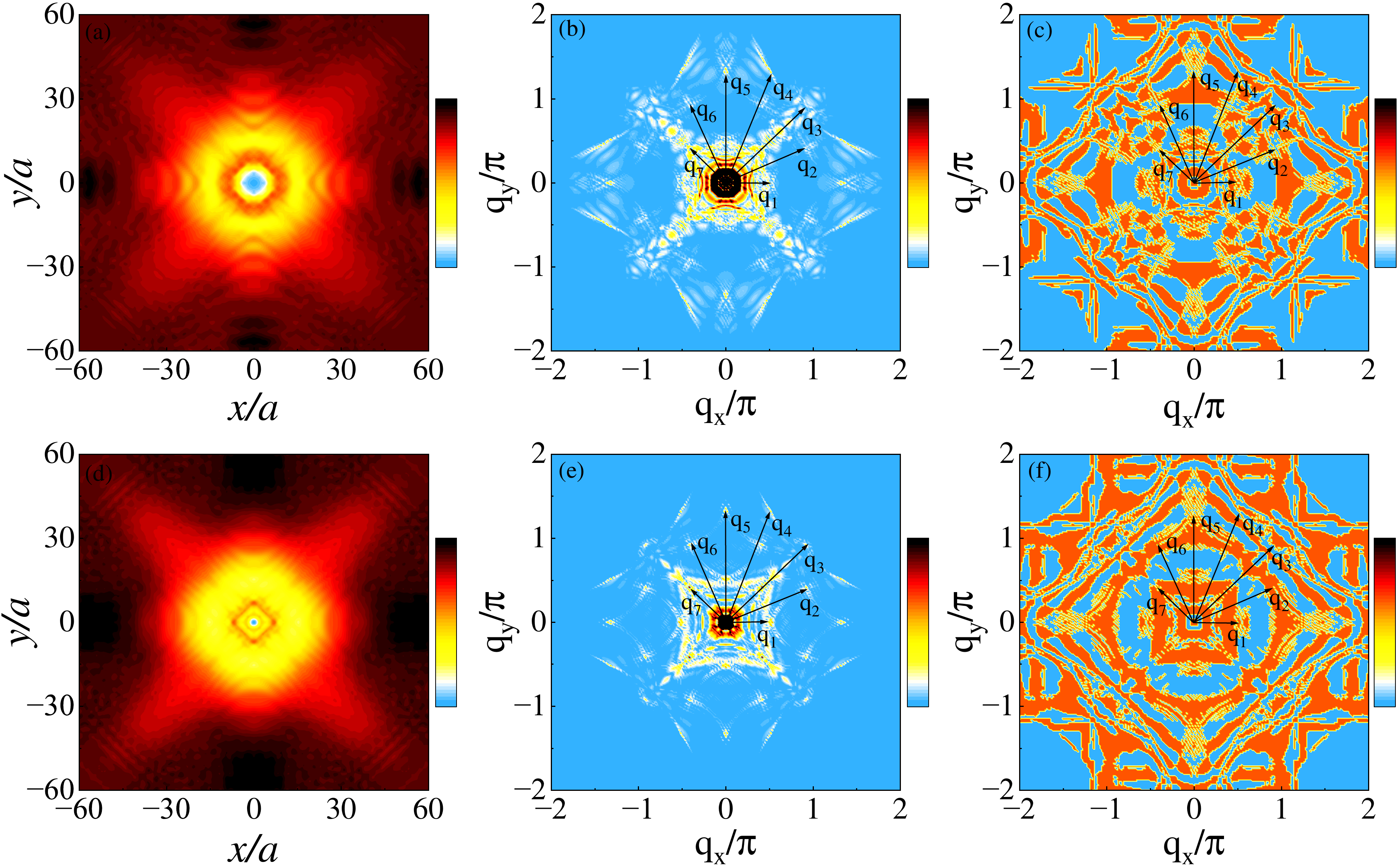}
\caption{(Color online) Upper panel: the maps of (a) the filtered local density of
states in real-space, (b) the amplitude of the filtered local density of states
in momentum-space, and (c) the phase of the filtered local density of states for an
out-of-plane single impurity in $\omega=-0.16J$ at $\delta=0.15$ with $T=0.002J$
for the impurity-scattering strength $V_{s}=5J$ and the screening length $L=5a$.
Lower panel: the corresponding maps of (d) the local density of states in
real-space, (e) the amplitude of the local density of states in momentum-space, and
(f) the phase of the local density of states in the case of the absence of the
filter effect for an out-of-plane single impurity. \label{LDOS-bare-filtered-out}}
\end{figure*}
In particular, in
correspondence to the enhancement of the filtered LDOS intensity along the
parallel direction, the intensities of the QSI peaks along
the parallel direction become more pronounced. Moreover, the central region around
the zero wave vector with the highest intensity is enlarged; (iii) the phase
distribution of the filtered LDOS exhibits an overall $C_4$
rotational symmetry, which is limited to two values of $0$ and $\pi$ due to the
spatial inversion symmetry of system with a single impurity at the origin in
real-space. Moreover, the phase of the QSI peaks with the
wave vectors ${\bf q}_{1}$ and ${\bf q}_{5}$ differs from these with the wave vectors
${\bf q}_{2}$ and ${\bf q}_{6}$ by the phase $\pi$, and thus exhibits a local d-wave
symmetry around the parallel direction. The above results in
Fig. \ref{LDOS-bare-filtered-in} thus show clearly that although there are some
subtle differences, the main features, such as the amplitude of LDOS in
momentum-space with the locations of the QSI peaks, remain unchanged even in the
presence of the {\it filter effect}\cite{Fujita14,Torre16NP,Kreisel15}.

For a further understanding of the nature of the filter effect on the LDOS
modulation, we have also performed a calculation for the filtered LDOS in the case
of the presence of an out-of-plane single impurity, and the results of (a) the
filtered LDOS in real-space, (b) the amplitude of the filtered LDOS in
momentum-space, and (c) the
phase of the filtered LDOS for an out-of-plane single impurity in $\omega=-0.16J$
at $\delta=0.15$ with $T=0.002J$ for the impurity-scattering strength $V_{s}=5J$
and the screening length $L=5a$ are mapped in the upper panel of
Fig. \ref{LDOS-bare-filtered-out} in comparison with the corresponding results of
(d) LDOS in real-space, (e) the amplitude of LDOS in momentum-space, and (f) the
phase of LDOS in the case of the absence of the filter effect for an out-of-plane
single impurity, respectively, in the lower panel, where although the overall
feature of the LDOS modulations for an in-plane single impurity and an out-of-plane
single impurity is similar, several substantial differences emerge: (i) the
suppression of the intensity of the filtered LDOS in real-space along the diagonal
direction from the d-wave factor in the filter function (\ref{d-wave-factor}) is
weaker than that for an in-plane single impurity as shown in
Fig. \ref{LDOS-bare-filtered-in}a, while the intensity in the central region around
the impurity site is heavily suppressed; (ii) the square-shape pattern of the LDOS
phase in momentum-space around the central region in the case of the absence of
the filter effect is transformed into the diamond-shape pattern due to the presence
of filter effect; (iii) the phase distribution pattern of the filtered LDOS has
a nodal and antinodal dichotomy, i.e., the phase distribution pattern along the
parallel direction displays a regular pattern similar to the case for an in-plane
single impurity, while the phase distribution pattern along the diagonal
direction behaves very differently and exhibits a random pattern, which leads to
the phase distribution in large momentums very fragmented and irregular. Moreover,
although the phase distribution pattern also exhibits a $C_{4}$ rotational symmetry
similar to the case for an in-plane single impurity, the phase distribution along
the parallel direction displays a local d-wave behaviour, which show clearly that
the d-wave factor in the filter function (\ref{d-wave-factor}) leads to the d-wave
pattern of the phase distribution along parallel direction, and therefore highlights
the role played by the filter effect.

Finally, it should be emphasized that the results of the broadened QSI peaks in the
filtered LDOS in momentum-space in Fig. \ref{LDOS-bare-filtered-in} and
Fig. \ref{LDOS-bare-filtered-out} are inconsistent with the corresponding
experimental observations
\cite{Hussey02,Balatsky06,Alloul09,Pan00,Hoffman02,McElroy03,Hanaguri07,Kohsaka08,Hanaguri09,Lee09,Vishik09,Schmidt11,Fujita14},
i.e., the results obtained from the filter function (\ref{d-wave-factor}) are
difficult to explain the sharp QSI peaks and the local d-wave distribution of the
LDOS phase along the parallel direction simultaneously, which calls for further
studies with more proper {\it filter function} to model the LDOS modulation observed
from STM/S experiments
\cite{Hussey02,Balatsky06,Alloul09,Pan00,Hoffman02,McElroy03,Hanaguri07,Kohsaka08,Hanaguri09,Lee09,Vishik09,Schmidt11,Fujita14}.

\subsection{Multiple impurities}

We now focus on the LDOS modulation in the case for a finite impurity concentration.
The above calculation of the $T$-matrix for a single impurity in
Eq. (\ref{Tmat-Expression}) [see Appendix \ref{matrix}] can be generalized
straightforwardly to the case for multiple noninteracting impurities. In the presence
of multiple impurities located at positions $\bm{R}_{i}$, the impurity scattering
potential for the electron at $\bm{r}$ can be expressed as
$V(\bm{r})=\sum_{i}V_{\rm I}(\bm{r}-\bm{R}_{i})$, while its form in momentum-space
can be derived via the Fourier-transform as,
\begin{widetext}
\begin{eqnarray}\label{VPOT-3}
V_{\bm{k}\bm{k}'} = \int d{\bf r}e^{-i(\bm{k}-\bm{k}')\cdot\bm{r}}
\sum_{i} V_{\rm I}(\bm{r}-\bm{R}_{i})
= \sum_{i}e^{-i(\bm{k}-\bm{k}')\cdot\bm{R}_{i}}\int d{\bf r}e^{-i(\bm{k}-\bm{k}')
\cdot\bm{r}}V_{\rm I}(\bm{r})
= V_{\bm{k}\bm{k}'}\sum_{i}e^{-i(\bm{k}-\bm{k}')\cdot\bm{R}_{i}}.
\end{eqnarray}
With the help of the above form of the multiple impurity scattering potential in
Eq. (\ref{VPOT-3}) and the in-plane and out-of-plane single impurity scattering
potentials in Eqs. (\ref{VPOT-RS}) and (\ref{VPOT-2-1}), the impurity scattering
potentials for the electron located at $\bm{r}$ with finite in-plane and
out-of-plane impurity concentrations can be expressed in real-space as,
\begin{eqnarray}
V(\bm{r})=\left
\{\begin{array}{ll}
\sum_{i}{V_{s}\over |\bm{r}-\bm{R}_{i}|}e^{-|\bm{r}-\bm{R}_{i}|/L},
~~{\rm for~in~plane},\\
\sum_{i}V_{s}e^{-|\bm{r}-\bm{R}_{i}|/L},~~{\rm for~out~of~plane},\\
\end{array}\right.
\end{eqnarray}
respectively, while their forms in momentum-space can be derived directly in terms
of the Fourier-transformation as,
\begin{eqnarray}\label{ISP-FIN}
V_{\bm{k}\bm{k}'}=\left
\{\begin{array}{ll}
{2\pi V_{s}\over \sqrt{({\bf k}-{\bf k}')^{2}+1/L^2}}\sum_{i}
e^{-i({\bf k}-{\bf k}')\cdot\bm{R}_{i}}, ~~{\rm for~in~plane},\\
{2\pi V_{s}/L\over [({\bf k}-{\bf k}')^{2}+1/L^{2}]^{3/2}}\sum_{i}
e^{-i({\bf k}-{\bf k}')\cdot\bm{R}_{i}}, ~~{\rm for~out~of~plane}.\\
\end{array}\right.
\end{eqnarray}
\end{widetext}

\begin{figure*}[t!]
\centering
\includegraphics[scale=0.20]{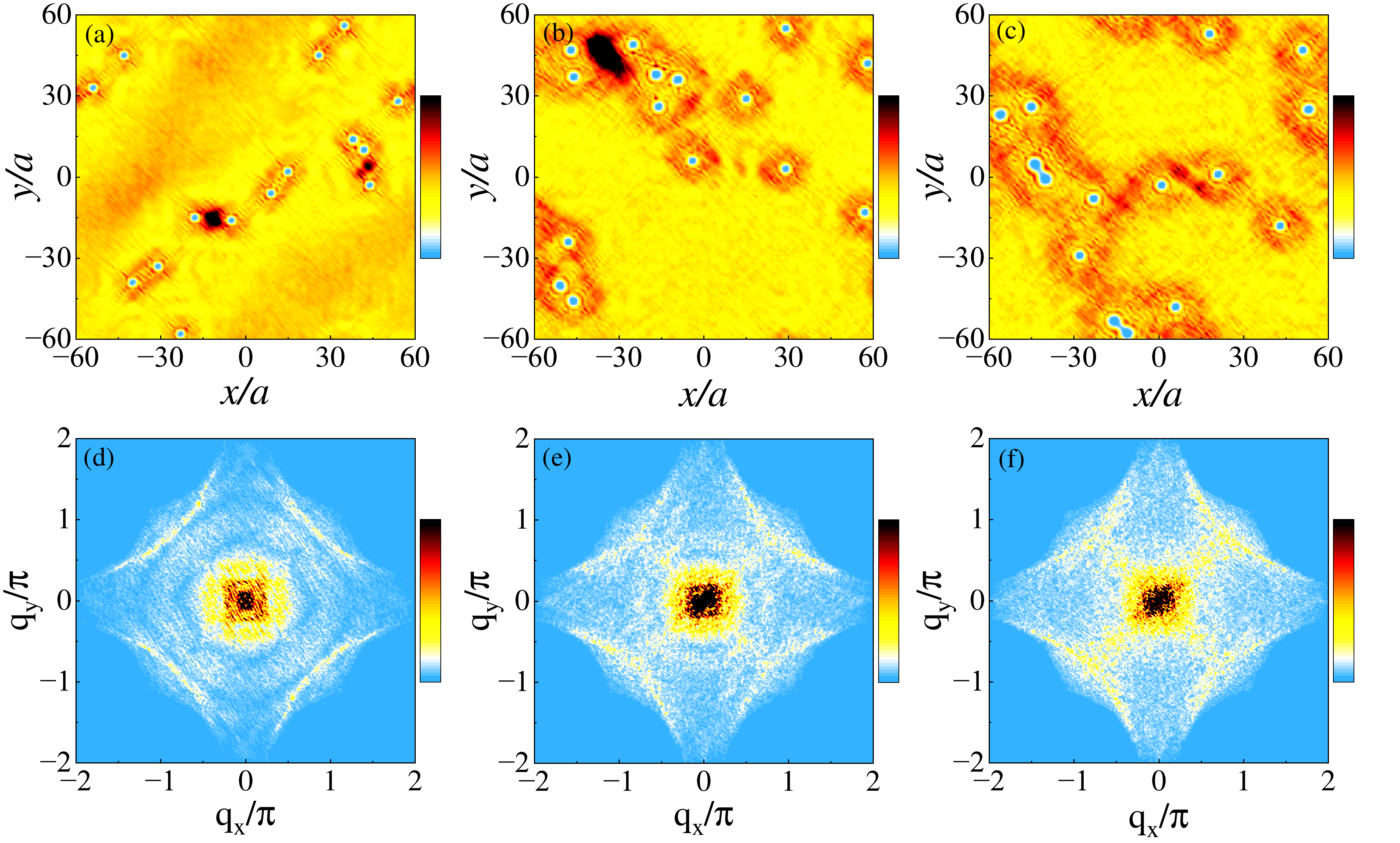}
\caption{(Color online) Upper panel: the maps of the local density of states in
real-space in the in-plane impurity concentration $n_{i}\approx 0.001$ and
$\omega=-0.16J$ at $\delta=0.15$ with $T=0.002J$ for the impurity-scattering
strength $V_{s}=8J$ and the screening lengths (a) $L=5a$, (b) $L=10a$, and (c)
$L=14a$. Lower panel: the corresponding maps of the amplitudes of the local density
of states in momentum-space for the screening lengths (d) $L=5a$, (e) $L=10a$, and
(f) $L=14a$ Fourier-transformed from the maps of the real-space local density of
states in (a), (b), and (c), respectively. \label{LDOS-RMS-L}}
\end{figure*}
\begin{figure*}[t!]
\centering
\includegraphics[scale=0.20]{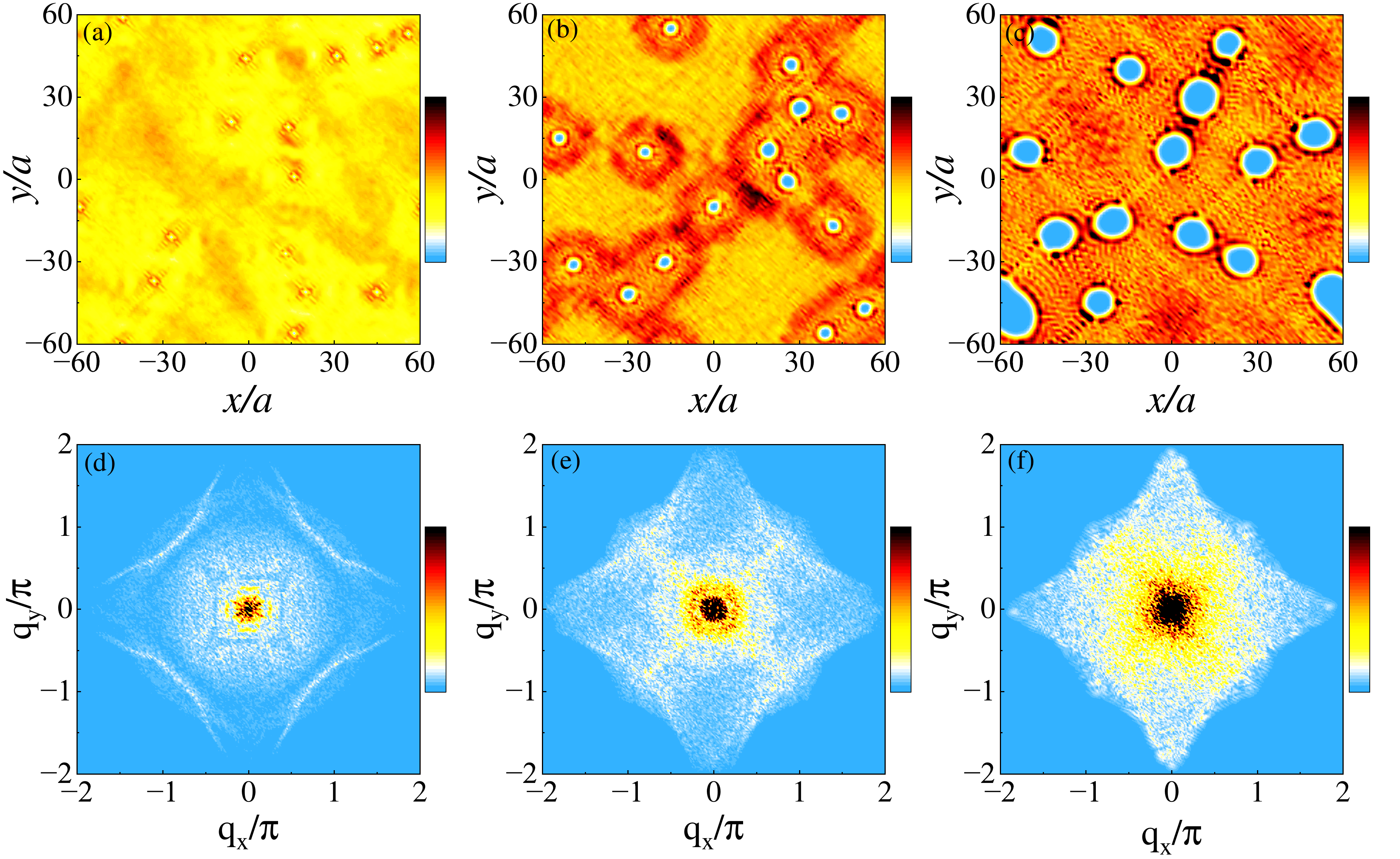}
\caption{(Color online) Upper panel: the maps of the local density of states in
real-space in the out-of-plane impurity concentration $n_{i}\approx 0.001$ and
$\omega=-0.16J$ at $\delta=0.15$ with $T=0.002J$ for the impurity-scattering
strength $V_{s}=5J$ and the screening lengths (a) $L=2a$, (b) $L=5a$, and (c)
$L=8a$. Lower panel: the corresponding maps of the amplitudes of the local density
of states in momentum-space for the screening lengths (d) $L=2a$, (e) $L=5a$, and
(f) $L=8a$ Fourier-transformed from the maps of the real-space local density of
states in (a), (b), and (c), respectively.\label{LDOS-RMS-L-B}}
\end{figure*}

Substituting the above multiple impurity scattering potential (\ref{ISP-FIN})
into Eqs. (\ref{Tmat-Expression}) and (\ref{Green-Fun-k-1}), LDOS in real-space
in Eq. (\ref{Rhoq-R}) and the Fourier transformed LDOS in momentum-space in
Eq. (\ref{Rhoq}) for finite in-plane and out-of-plane impurity concentrations
can be obtained explicitly. We are now ready to discuss the quasiparticle
scattering from multiple impurities. In Fig. \ref{LDOS-RMS-L}, we plot the LDOS
maps in real-space in the in-plane impurity concentration $n_{i}\approx 0.001$
and $\omega=-0.16J$ at $\delta=0.15$ with $T=0.002J$ for the impurity-scattering
strength $V_{s}=8J$ and the screening lengths (a) $L=5a$, (b) $L=10a$, and (c)
$L=14a$ in the upper panel,
where the impurity number is set as $N_{i} = 15$ for the finite $120\times 120$
lattice, while the impurity locations are determined by random sampling, and thus
are distributed randomly in real space.
In the lower panel, we plot the corresponding maps of the amplitudes of LDOS in
momentum-space for the screening lengths (d) $L=5a$, (e) $L=10a$, and (f) $L=14a$,
which are obtained from the real-space LDOS maps in (a), (b), and (c),
respectively, in terms of the Fourier-transform. For a clear comparison, the LDOS
maps in real-space in the out-of-plane impurity concentration $n_{i}\approx 0.001$
and $\omega=-0.16J$ at $\delta=0.15$ with $T=0.002J$ for the impurity-scattering
strength $V_{s}=5J$ and the screening lengths (a) $L=2a$, (b) $L=5a$, and (c)
$L=8a$ are plotted in the upper panel of Fig. \ref{LDOS-RMS-L-B}, while the
corresponding maps of the amplitudes of LDOS in momentum-space for the screening
lengths (d) $L=2a$, (e) $L=5a$, and (f) $L=8a$ Fourier-transformed from the maps
of the real-space LDOS in (a), (b), and (c), respectively are plotted in the lower
panel of Fig. \ref{LDOS-RMS-L-B}. The above results in Fig. \ref{LDOS-RMS-L} and
Fig. \ref{LDOS-RMS-L-B} thus show that the overall features of the LDOS modulation
pattern in momentum-space appear for a low in-plane (out-of-plane) impurity
concentration to be similar to that for an in-plane (out-of-plane) single impurity
shown in Fig. \ref{LDOS-IPSI}b (Fig. \ref{LDOS-OPSI}b), where (i) the LDOS
modulation for a low out-of-plane impurity concentration is more dramatic than that
for a low in-plane impurity concentration;
(ii) the intensity of the LDOS modulation in momentum-space is dominated by large
momenta, especially these along the diagonal direction, while the intensity around
the central region is dominated by small momenta. However, with the increase of the
screening length, the intensity from small momenta is enhanced, while the impurity
weight in Eq. \ref{imp-weight} from the large momenta is decreased, which are in a
qualitative agreement with a single impurity system. Moreover, the LDOS modulation
in momentum-space becomes more dispersive, and the features of the LDOS modulation
with short screening lengths are quite similar to the case for a single impurity
system, because the effects from the multiple-impurity scattering would be weakened
in short screening lengths.
However, although momentum-space structure of the LDOS modulation
pattern for a low impurity concentration is quite similar to the corresponding one
for a single impurity, some subtle differences between a single impurity and
multiple impurities emerge, where (i) the multiple impurities in
real-space distribute randomly. In particular, the impurity-scattering area (the
blue circles in the upper panel of Fig. \ref{LDOS-RMS-L-B}) for out-of-plane
multiple impurities increases with the increase of the screening length, which
indicates the gradual formation of the electronic puddles due to the scattering
interference in the system with a low out-of-plane impurity concentration when
the screening length is increased; (ii) the impurity region in real-space can be
divided into two parts: the isolate impurity region and the impurity region where
impurities get close together. The latter impurity region is induced due to the
multiple impurity scattering effects\cite{Zhu03a,Atkinson03a}. When two impurities
line up such that their diagonal streaks overlap each other, the streaks
constructively interfere and have the effect that they become intense
\cite{Sulangi17}; (iii) although the pronounced QSI peaks in the momentum-space
LDOS modulation pattern that are prominent for a single impurity as shown in
Fig. \ref{LDOS-IPSI}b and Fig. \ref{LDOS-OPSI}b are smeared heavily in the case for
a finite impurity concentration, they are similarity visible; (iv) the LDOS maps in
momentum-space for a low in-plane or a low out-of-plane impurity concentration
show a speckle pattern, which are very similar to the speckle textures seen in the
experiments
\cite{Hussey02,Balatsky06,Alloul09,Pan00,Hoffman02,McElroy03,Hanaguri07,Kohsaka08,Hanaguri09,Lee09,Vishik09,Schmidt11,Fujita14}.
In summary, except for some subtle differences between a single impurity and
multiple impurities, the outcome in the case for a low impurity concentration is a
speckled version of the single-impurity result.

\subsection{Spatially random on-site impurity}

Finally, we discuss the LDOS modulation for spatially random on-site impurity$-$a
random distribution of on-site energies throughout the sample, where the on-site
energies vary from one site to another, and then as a result, the numerous
multiple-scattering processes occur. In this case, the impurity-scattering potential
is discretized as $V(\bm{R}_{i})\in\{E_{1},E_{2},E_{3},\cdots,E_{N}\}$.
For simplicity, we will take $V(\bm{R}_{i})\in\{E_{1},E_{2},E_{3},\cdots, E_{N}\}$
to be drawn from a Gaussian distribution, and then the number of lattice site
$N_i$ with its impurity-scattering potential $E_{i}$ is determined from the
Gaussian-distribution as,
\begin{equation}
N_{i}={N\over \sqrt{\sigma\pi}}\int_{E_{i}-\Delta}^{E_{i}+\Delta}e^{-x^{2}/\sigma}
dx,
\end{equation}
where $N$ is the number of the lattice sites, $\Delta=(E_{i+1}-E_{i})/2$, the width
of the distribution is parameterized by the standard deviation $\sigma$, which
characterizes the strength of the impurity-scattering potential. The lattice site
on which the impurity-scattering potential is $E_{i}$ is determined
by using the random samples. Repeating above calculation until all
impurity-scattering potentials are assigned to all lattice sites.
\begin{figure}[h!]
\centering
\includegraphics[scale=0.40]{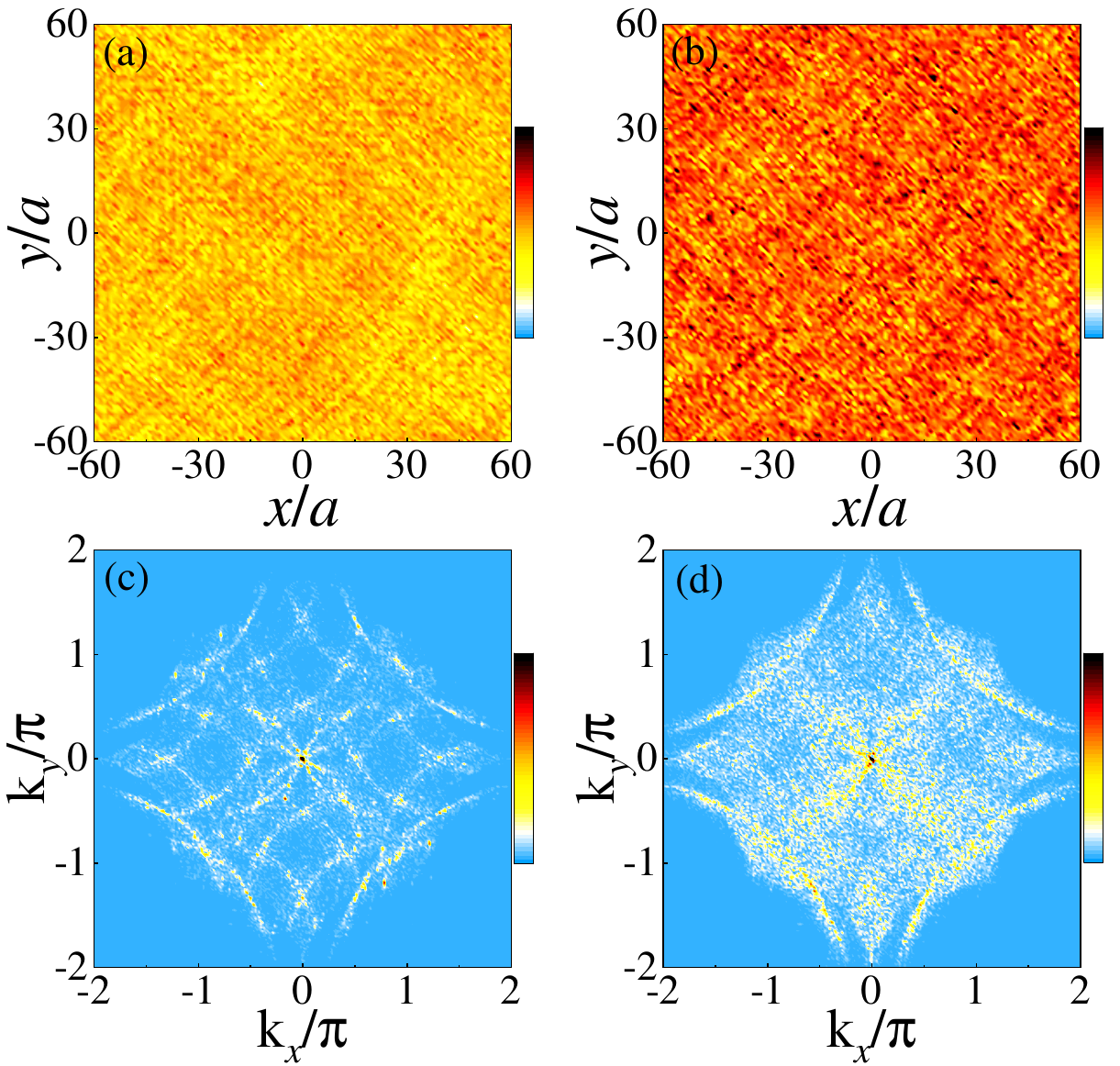}
\caption{(Color online) Upper panel: the maps of the local density of states in
real-space for Gaussian-random-distributed impurity in $\omega=-0.16$ at
$\delta=0.15$ with $T=0.002J$ and the standard deviations (a) $\sigma=0.05$ and
(b) $\sigma=0.1$. Lower panel: the corresponding maps of the local density of
states in momentum-space for the standard deviations (c) $\sigma=0.05$ and (d)
$\sigma=0.1$ Fourier-transformed from the real-space local density of states in
(a) and (b), respectively.\label{LDOS-RMS}}
\end{figure}
The form of this
impurity-scattering potential from Gaussian-random-distributed impurity in
momentum-space then is obtained directly via the Fourier transform, and can be
expressed explicitly as,
\begin{equation}
V_{\bm{k}\bm{k}'}=\sum_{l}V(\bm{R}_{l})e^{-i(\bm{k}-\bm{k}')\cdot\bm{R}_{l}}.
\end{equation}
It has been shown that this form of the impurity could potentially feature both
small and large scattering wave vector peaks in the LDOS modulation spectrum
\cite{Sulangi17,Atkinson00a}.

In Fig. \ref{LDOS-RMS}, we plot the LDOS maps in real-space for
Gaussian-random-distribution of on-site impurity in $\omega=-0.16$ at $\delta=0.15$
with $T=0.002J$ and the standard deviations (a) $\sigma=0.05$ and (b) $\sigma=0.1$
in the upper panel. In the lower panel, we plot the corresponding LDOS maps in
momentum-space for the standard deviations (c) $\sigma=0.05$ and (d) $\sigma=0.1$
Fourier-transformed from the real-space LDOS in (a) and (b), respectively, where
the main features of the LDOS modulation can be summarized as: (i) the LDOS maps in
real-space is characterized by a number of isolate islands with relatively higher
intensities, however, the arrangements of these isolate islands seem not to be a
similar extent as in the case for an in-plane single impurity or an out-of-plane
single impurity; (ii) with the increase of the standard deviation, although the
number of isolate islands in real-space increases, their distribution tends to be
more disordered; (iii) in a striking analogy to the case for multiple impurities,
the randomness of the positions is that the LDOS maps in momentum-space show speckle
patterns. In particular, the momentum-space structure of the LDOS modulation at a
relatively large deviation displays a crisscrossing (in the diagonal directions)
pattern slightly similar to that obtained for multiple impurities, although the
crisscrossing patterns are far more subdued. The above results thus show that the
single-impurity effect in the momentum-space LDOS modulation is partially mitigated
by Gaussian on-site potential impurity, and then the momentum-space LDOS modulation
for Gaussian-random-distribution of on-site impurity at a relatively large deviation
displays a similar behavior of the momentum-space LDOS modulation for multiple
impurities.

\section{Summary and discussion}\label{conclude}

In the framework of the kinetic-energy-driven superconductivity, we have rederived
the homogeneous part of the electron propagator. This homogeneous electron
propagator produces a constant energy contour with the most of the spectral weight
located at around the tips of the Fermi arcs, and then these tips of the Fermi arcs
connected by the quasiparticle scattering wave vectors ${\bf q}_{i}$
naturally construct an {\it octet scattering model}. Starting from this homogeneous
electron propagator and the related microscopic octet scattering model, we have
studied the nature of the LDOS modulation in cuprate superconductors within the
$T$-matrix approach, where we have developed a new method of the inversion of matrix
to accurately derive the $T$-matrix generated by the quasiparticle scattering from
various kinds of impurities. The obtained $T$-matrix then are employed to derive
LDOS firstly by the involvement of all the quasiparticle excitations and scattering
processes. Our results show that although there are some subtle differences between
the LDOS modulations for an in-plane single impurity and an out-of-plane single
impurity, the overall features of the LDOS modulation in cuprate superconductors
can be qualitatively explained by taking into account the quasiparticle scattering
from a single impurity on the kinetic-energy-driven homogeneous SC-state, where the
pronounced QSI peaks are located at the well-defined scattering wave vectors
${\bf q}_{i}$, while these QSI scattering wave vectors ${\bf q}_{i}$ and the related
QSI peak dispersions are internally consistent within the octet scattering model.
However, these pronounced QSI peaks in the momentum-space LDOS modulation pattern
for a single impurity are smeared heavily in the case for multiple impurities.
Moreover, due to the quasiparticle scattering effects from multiple impurities, the
momentum-space LDOS modulation for a low impurity concentration exhibits a speckle
pattern. In other words, the outcome in the case for a low-impurity concentration is
a speckled version of the single-impurity result.
Moreover, the main features, such as the amplitude of LDOS in momentum-space with
the locations of the QSI peaks, remain unchanged even in the presence of the filter
effect except for that the filter effect induces a local d-wave symmetry around the
parallel direction of the LDOS phase in momentum-space.
On the other hand, the
momentum-space LDOS modulation for Gaussian-random-distribution of on-site impurity
at a relatively large deviation displays a similar behavior of the momentum-space
LDOS modulation for multiple impurities. Our theory also indicates that the impurity
weight almost linearly increases with the increase of the impurity scattering
strength for the weak scattering strength and saturates to a fixed value for the
strong scattering strength, while it decreases with the increase of the impurity
scattering screening length for the short screening length and saturates for the
long screening length.

A natural question is what is the reason why the present theory can give a
consistent description of QSI in cuprate superconductors? To our present
understanding, there are at least three reasons: (i) the strong electron
correlation in the homogeneous part of the electron propagator has been treated
properly based on the kinetic-energy-driven superconductivity, as a consequence,
the results of the unconventional electronic structure\cite{Zeng22,Gao19} are
qualitatively consistent with the corresponding ARPES experimental observations;
(ii) {\it the microscopic octet scattering model} obtained in the homogeneous
system can persist into the case for the presence of the impurity scattering
\cite{Zeng22,Gao19}, while this octet scattering model is a basic scattering model
in the explanation of the Fourier transform STM/S experimental data
\cite{Hussey02,Balatsky06,Alloul09,Pan00,Hoffman02,McElroy03,Hanaguri07,Kohsaka08,Hanaguri09,Lee09,Vishik09,Schmidt11,Fujita14};
(iii) the $T$-matrix obtained accurately in terms of the new method of the
inversion of matrix is essential,
since the accurate calculation of the $T$-matrix for various kinds of impurities
contains all the quasiparticle excitations and scattering processes. This is why
the present theory can be used to discuss the impurity scattering from the real
impurity potentials of the system. In this case, the difficulties appeared in some
other traditional methods, such as the perturbation theory in the Born limit, the
traditional self-consistent $T$-matrix approach, and the Bogoliubov-de Gennes
equations, for the discussions of the impurity scattering have been overcome
in the present theory. Furthermore, the theory developed in this paper for the
understanding of QSI in the strongly electron correlated square-lattice cuprate
superconductors can be also employed to study QSI in other interesting systems,
such as the strongly electron correlated honeycomb-lattice, triangular-lattice, and
kagome-lattice superconductors. These and the related issues are under investigation
now.

\begin{acknowledgments}

This work is supported by the National Key Research and Development Program of China
under Grant Nos. 2023YFA1406500 and 2021YFA1401803, and the National Natural Science
Foundation of China under Grant Nos. 12247116 and 12274036.

\end{acknowledgments}

\appendix

\section{T-matrix}\label{matrix}

In this Appendix, our goal is to obtain the $T$-matrix
$\tilde{T}_{\bm{k}\bm{k}'}(\omega)$ in Eq. (\ref{T-matrix-equation-1}) of the main
text. To obtain this $T$-matrix $\tilde{T}_{\bm{k}\bm{k}'}(\omega)$, we right
multiply the matrix $\tau_3$ in Eq. (\ref{T-matrix-equation-1}) of the main text,
and then rewrite it as,
\begin{eqnarray}\label{Tmat-Eq}
&&\tilde{T}_{\bm{k}\bm{k}'}(\omega)\tau_{3}={1\over N}V_{\bm{k}\bm{k}'}\tau_{0}
+{1\over N}\sum_{\bm{k}_{1}}V_{\bm{k}\bm{k}_{1}}\tau_{3}\tilde{G}(\bm{k}_{1},\omega)
{1\over N}V_{\bm{k}_{1}\bm{k}'}\nonumber\\
&&+{1\over N}\sum_{\bm{k}_{1}}V_{\bm{k}\bm{k}_{1}}\tau_{3}\tilde{G}(\bm{k}_{1},\omega)
{1\over N}\sum_{\bm{k}_{2}}V_{\bm{k}_{1}\bm{k}_{2}}\tau_{3}
\tilde{G}(\bm{k}_{2},\omega){1\over N}V_{\bm{k}_{2}\bm{k}'}\nonumber\\
&&+ \cdots.
\end{eqnarray}
On the other hand, the $T$-matrix $\tilde{T}_{\bm{k}\bm{k}'}(\omega)$ can be also
separated in terms of its components $T^{\mu}_{\bm{k}\bm{k}'}(\omega)$
into $\tilde{T}_{\bm{k}\bm{k}'}(\omega)=\sum_{\mu}T^{\mu}_{\bm{k}\bm{k}'}(\omega)
\tau_{\mu}$, while these components can be obtained as\cite{Zeng22},
\begin{equation}\label{Tmat-Solution}
\sum\limits_{\mu}T^{\mu}_{\bm{k}\bm{k'}}(\omega)\tau_{\mu}\tau_{3}
=\left ( \bar{V}\otimes\tau_{0}{1\over 1-\bar{M}}\right )_{\bm{k}\bm{k'}},
\end{equation}
where the matrices $\bar{V}$ and $\bar{M}$ can be derived explicitly as
\cite{Zeng22},
\begin{widetext}
\begin{subequations}
\begin{eqnarray}
\bar{V}\otimes\tau_{0}&=&{1\over N}V\otimes\tau_{0}={1\over N}
\left(\begin{array}{ccccc}
V_{\bm{k}_{1}\bm{k}_{1}}\tau_{0}& V_{\bm{k}_{1}\bm{k}_{2}}\tau_{0}&
V_{\bm{k}_{1}\bm{k}_{3}}\tau_{0}& \cdots & V_{\bm{k}_{1}\bm{k}_N}\tau_{0}\\
V_{\bm{k}_{2}\bm{k}_{1}}\tau_{0}& V_{\bm{k}_{2}\bm{k}_{2}}\tau_{0}&
V_{\bm{k}_{2}\bm{k}_{3}}\tau_{0}& \cdots & V_{\bm{k}_{2}\bm{k}_N}\tau_{0}\\
\vdots & \vdots & \vdots & \vdots & \vdots\\
V_{\bm{k}_N\bm{k}_{1}}\tau_{0}& V_{\bm{k}_{N}\bm{k}_{2}}\tau_{0}&
V_{\bm{k}_{N}\bm{k}_{3}}\tau_{0}& \cdots & V_{\bm{k}_{N}\bm{k}_{N}}\tau_{0}
\end{array}\right),\\
\bar{M}&=&{1\over N}
\left(\begin{array}{ccccc}
\tau_{3}\tilde{G}(\bm{k}_{1},\omega)V_{\bm{k}_{1}\bm{k}_{1}} &
\tau_{3}\tilde{G}(\bm{k}_{1},\omega)V_{\bm{k}_{1}\bm{k}_{2}} &
\tau_{3}\tilde{G}(\bm{k}_{1},\omega)V_{\bm{k}_{1}\bm{k}_{3}} & \cdots &
\tau_{3}\tilde{G}(\bm{k}_{1},\omega)V_{\bm{k}_{1}\bm{k}_{N}} \\
\tau_{3}\tilde{G}(\bm{k}_{2},\omega)V_{\bm{k}_{2}\bm{k}_{1}} &
\tau_{3}\tilde{G}(\bm{k}_{2},\omega)V_{\bm{k}_{2}\bm{k}_{2}} &
\tau_{3}\tilde{G}(\bm{k}_{2},\omega)V_{\bm{k}_{2}\bm{k}_{3}} & \cdots &
\tau_{3}\tilde{G}(\bm{k}_{2},\omega)V_{\bm{k}_{2}\bm{k}_{N}} \\
\vdots & \vdots & \vdots & \vdots & \vdots\\
\tau_{3}\tilde{G}(\bm{k}_{N},\omega)V_{\bm{k}_{N}\bm{k}_{1}} &
\tau_{3}\tilde{G}(\bm{k}_{N},\omega)V_{\bm{k}_{N}\bm{k}_{2}} &
\tau_{3}\tilde{G}(\bm{k}_{N},\omega)V_{\bm{k}_{N}\bm{k}_{3}} & \cdots &
\tau_{3}\tilde{G}(\bm{k}_{N},\omega)V_{\bm{k}_{N}\bm{k}_{N}}
\end{array}\right),
\end{eqnarray}
\end{subequations}
respectively, and then the $T$-matrix $\tilde{T}_{\bm{k}\bm{k}'}(\omega)$ can be
obtained directly from Eq. (\ref{Tmat-Solution}) as,
\begin{equation}
\tilde{T}(\omega)=\bar{V}\otimes\tau_{0}{1\over 1-\bar{M}}\hat{I}_{v}\otimes
\tau_{3},
\end{equation}
which is the same as quoted in Eq. (\ref{T-matrix-equation-1}) of the main text.

\end{widetext}


\begin{thebibliography}{00}

\bibitem{Carbotte11} See e.g., the review, J. P. Carbotte, T. Timusk, and J. Hwang,
Bosons in high-temperature superconductors: an experimental survey,
Rep. Prog. Phys. {\bf 74}, 066501 (2011).

\bibitem{Bok16} J. M. Bok, J. J. Bae, H.-Y. Choi, C. M. Varma, W. Zhang, J. He, Y.
Zhang, L. Yu, and X. J. Zhou, Quantitative determination of pairing interactions
for high-temperature superconductivity in cuprates, Sci. Adv. {\bf 2}, e1501329
(2016).

\bibitem{Damascelli03} See, e.g., the review, A. Damascelli, Z. Hussain, and Z.-X.
Shen, Angle-resolved photoemission studies of the cuprate superconductors, Rev.
Mod. Phys. {\bf 75}, 473 (2003).

\bibitem{Campuzano04} See, e.g., the review, J. C. Campuzano, M. R. Norman, M.
Randeira, Photoemission in the high-$T_c$ superconductors, in {\it Physics of
Superconductors}, vol. II, edited by K. H. Bennemann and J. B. Ketterson (Springer,
Berlin Heidelberg New York, 2004), p. 167.

\bibitem{Fischer07} See, e.g., the review, \O . Fischer, M. Kugler, I.
Maggio-Aprile, C. Berthod, and C. Renner, Scanning tunneling spectroscopy of
high-temperature superconductors, Rev. Mod. Phys. {\bf 79}, 353 (2007).

\bibitem{Yin21} J.-X. Yin, S. H. Pan, M. Z. Hasan, Probing topological quantum matter
with scanning tunnelling microscopy, Nat. Rev. Phys. {\bf 3}, 249 (2021).

\bibitem{Bednorz86} J. G. Bednorz and K. A. M\"uller, Possible high $T_{c}$
superconductivity in the Ba-La-Cu-O system, Z. Phys. B {\bf 64}, 189 (1986).

\bibitem{Wu87} M. K. Wu, J. R. Ashburn, C. J. Torng, P. H. Hor, R. L. Meng, L. Gao,
Z. J. Huang, Y. Q. Wang, and C. W. Chu, Superconductivity at 93 K in a new
mixed-phase Y-Ba-Cu-O compound system at ambient pressure, Phys. Rev. Lett.
{\bf 58}, 908 (1987).

\bibitem{Anderson87} P. W. Anderson, The resonating valence bond state in
La$_{2}$CuO$_{4}$ and superconductivity, Science {\bf 235}, 1196 (1987).

\bibitem{Comin16} See, e.g., the review, R. Comin and A. Damascelli, Resonant X-ray
scattering studies of charge order in cuprates, Annu. Rev. Condens. Matter Phys.
{\bf 7}, 369 (2016).

\bibitem{Vishik18} See, e.g., the review, I. M. Vishik, Photoemission perspective
on pseudogap, superconducting fluctuations, and charge order in cuprates:
a review of recent progress, Rep. Prog. Phys. {\bf 81},062501 (2018).

\bibitem{Pan01} S. H. Pan, J. P. \'ONeal, R. L. Badzey, C. Chamon, H. Ding, J. R.
Engelbrecht, Z. Wang, H. Eisaki, S. Uchida, A. K. Gupta, K.-W. Ng, E. W. Hudson, K.
M. Lang, and J. C. Davis, Microscopic electronic inhomogeneity in the high-$T_{c}$
superconductor Bi$_{2}$Sr$_{2}$CaCu$_{2}$O$_{8+x}$, Nature {\bf 413}, 282 (2001).

\bibitem{Vershinin04} M. Vershinin, S. Misra, S. Ono, Y. Abe, Y. Ando, A. Yazdani,
Local ordering in the pseudogap state of the high-$T_{c}$ superconductor
Bi$_{2}$Sr$_{2}$CaCu$_{2}$O$_{8+\delta}$, Science {\bf 303}, 1995 (2004).

\bibitem{Hussey02} See, e.g., the review, N. E. Hussey, Low-energy quasiparticles
in high-$T_c$ cuprates,  Adv. Phys. {\bf 51}, 1685 (2002).

\bibitem{Balatsky06} See, e.g., the review, A. V. Balatsky, I. Vekhter, and J.-X.
Zhu, Impurity-induced states in conventional and unconventional superconductors,
Rev. Mod. Phys. {\bf 78}, 373 (2006).

\bibitem{Alloul09} See, e.g., the review, H. Alloul, J. Bobroff, M. Gabay, and P. J.
Hirschfeld, Defects in correlated metals and superconductors, Rev. Mod. Phys.
{\bf 81}, 45 (2009).

\bibitem{Pan00} S. H. Pan, E. W. Hudson, K. M. Lang, H. Eisaki, S. Uchida, and
J. C. Davis, Imaging the effects of individual zinc impurity atoms on
superconductivity in Bi$_{2}$Sr$_{2}$CaCu$_{2}$O$_{8+\delta}$, Nature {\bf 403},
746 (2000).

\bibitem{Hoffman02} J. E. Hoffman, E. W. Hudson, K. M. Lang, V. Madhavan, H. Eisaki,
S. Uchida, and J. C. Davis, Imaging quasiparticle interference in
Bi$_{2}$Sr$_{2}$CaCu$_{2}$O$_{8+\delta}$, Science {\bf 297}, 1148 (2002).

\bibitem{McElroy03} K. McElroy, R. W. Simmonds, J. E. Hoffman, D.-H. Lee, J.
Orenstein, H. Eisaki, S. Uchida, and J. C. Davis, Relating atomic-scale electronic
phenomena to wave-like quasiparticle states in superconducting
Bi$_{2}$Sr$_{2}$CaCu$_{2}$O$_{8+\delta}$, Nature {\bf 422}, 592 (2003).

\bibitem{Hanaguri07} T. Hanaguri, Y. Kohsaka, J. C. Davis, C. Lupien, I. Yamada, M.
Azuma, M. Takano, K. Ohishi, M. Ono, and H. Takagi, Quasiparticle interference and
superconducting gap in Ca$_{2-x}$Na$_{x}$CuO$_{2}$Cl$_{2}$, Nature Phys. {\bf 3},
865 (2007).

\bibitem{Kohsaka08} Y. Kohsaka, C. Taylor, P. Wahl, A. Schmidt, J. Lee, K. Fujita, J.
W. Alldredge, K. McElroy, J. Lee, H. Eisaki, S. Uchida, D.-H. Lee, and J. C. Davis,
How Cooper pairs vanish approaching the Mott insulator in
Bi$_{2}$Sr$_{2}$CaCu$_{2}$O$_{8+\delta}$, Nature {\bf 454}, 1072 (2008).

\bibitem{Hanaguri09} T. Hanaguri, Y. Kohsaka, M. Ono, M. Maltseva, P. Coleman, I.
Yamada, M. Azuma, M. Takano, K. Ohishi, and H. Takagi, Coherence factors in a
high-$T_{c}$ cuprate probed by quasi-particle scattering off vortices, Science
{\bf 323}, 923 (2009).

\bibitem{Lee09} J. Lee, K. Fujita, A. R. Schmidt, C. K. Kim, H. Eisaki, S. Uchida,
and J. C. Davis, Spectroscopic fingerprint of phase-incoherent superconductivity in
the underdoped Bi$_{2}$Sr$_{2}$CaCu$_{2}$O$_{8+\delta}$, Science {\bf 325}, 1099
(2009).

\bibitem{Vishik09} I. M. Vishik, E. A. Nowadnick, W. S. Lee, Z. X. Shen, B. Moritz,
T. P. Devereaux, K. Tanaka, T. Sasagawa, and T. Fujii, A momentum-dependent
perspective on quasiparticle interference in
Bi$_{2}$Sr$_{2}$CaCu$_{2}$O$_{8+\delta}$, Nat. Phys. {\bf 5}, 718 (2009).

\bibitem{Schmidt11} A. R. Schmidt, K. Fujita, E.-A. Kim, M. J. Lawler, H. Eisaki,
S. Uchida, D.-H. Lee, and J. C. Davis, Electronic structure of the cuprate
superconducting and pseudogap phases from spectroscopic imaging STM,
New J. Phys. {\bf 13}, 065014 (2011).

\bibitem{Fujita14} K. Fujita, M. H. Hamidian, S.D. Edkins, Chung Koo Kim, Y. Kohsaka,
M. Azuma, M. Takano, H. Takagi, H. Eisaki, S. Uchida, A. Allais, M. J. Lawler, E. -A.
Kim, S. Sachdev, and J. C. S\'eamus Davis, Direct phase-sensitive identification of a
$d$-form factor density wave in underdoped cuprates,
Proc. Natl Acad. Sci. USA {\bf 111}, E3026-E3032 (2014).

\bibitem{Capriotti03} L. Capriotti, D. J. Scalapino, and R. D. Sedgewick,
Wave-vector power spectrum of the local tunneling density of states: Ripples in a
$d$-wave sea, Phys. Rev. B {\bf 68}, 014508 (2003).

\bibitem{Wulindan10} D. Wulin, C.-C. Chien, D. K. Morr, and K. Levin,
Contrasting nodal and antinodal behavior in the cuprates via multiple gap
spectroscopies, Phys. Rev. B {\bf 81}, 100504(R) (2010).

\bibitem{Nowadnick12} E. A. Nowadnick, B. Moritz, and T. P. Devereaux,
Quasiparticle interference and the interplay between superconductivity and density
wave order in the cuprates, Phys. Rev. B {\bf 86}, 134509 (2012).

\bibitem{Torre16} E. D. Torre, D. Benjamin, Y. He, D. Dentelski, and E. Demler,
Friedel oscillations as a probe of fermionic quasiparticles,
Phys. Rev. B {\bf 93}, 205117 (2016).

\bibitem{Torre16NP} E. D. Torre, Y. He, and E. Demler, Holographic maps of
quasiparticle interference, Nat. Phys. {\bf 12}, 1052 (2016).

\bibitem{Wang03} Q.-H. Wang and D.-H. Lee, Quasiparticle scattering interference in
high-temperature superconductors, Phys. Rev. B {\bf 67}, 020511(2003).

\bibitem{Zhang03} D. Zhang and C. S. Ting, Energy-dependent modulations in the local
density of states of the cuprate superconductors, Phys. Rev. B {\bf 67}, 100506(R)
(2003).

\bibitem{Zhu04a} L. Zhu, W. A. Atkinson, and P. J. Hirschfeld, Power spectrum of
many impurities in a $d$-wave superconductor,
Phys. Rev. B {\bf 69}, 060503(R) (2004).

\bibitem{ZhaoMM23} M. Zhao, W.-W. Yang, H.-G. Luo, and Y. Zhong, Friedel oscillation
in non-Fermi liquid: Lesson from exactly solvable Hatsugai-Kohmoto model,
J. Phys.: Condens.Matter {\bf 35}, 495603 (2023).

\bibitem{Wangshuhua15} S.-H. Wang, H.-S. Zhao, and F. Yuan, Quasiparticle scattering
interference in the renormalized Hubbard model,
Front. Phys. {\bf 10}, 107401 (2015).

\bibitem {Kreisel15} A. Kreisel, P. Choubey, T. Berlijn, W. Ku, B. M. Andersen, and
P. J. Hirschfeld,
Interpretation of scanning tunneling quasiparticle interference and impurity states
in cuprates, Phys. Rev. Lett. {\bf 114}, 217002 (2015).

\bibitem {Choubey17} P. Choubey, A. Kreisel, T. Berlijn, B. M. Andersen, and P. J.
Hirschfeld, Universality of scanning tunneling microscopy in cuprate superconductors,
Phys. Rev. B {\bf 96}, 174523 (2017).

\bibitem{Sulangi17} M. A. Sulangi, M. P. Allan, and J. Zaanen, Revisiting
quasiparticle scattering interference in high-temperature superconductors: The
problem of narrow peaks, Phys. Rev. B {\bf 96}, 134507 (2017).

\bibitem{Sulangi18-a} M. A. Sulangi and J. Zaanen, Self-energies and quasiparticle
scattering interference, Phys. Rev. B {\bf 98}, 094518 (2018).

\bibitem{Sulangi18-b} M. A. Sulangi and J. Zaanen, Quasiparticle density of states,
localization, and distributed disorder in the cuprate superconductors,
Phys. Rev. B {\bf 97}, 144512 (2018).

\bibitem{Chatterjee06} U. Chatterjee, M. Shi, A. Kaminski, A. Kanigel, H. M.
Fretwell, K. Terashima, T. Takahashi, S. Rosenkranz, Z. Z. Li, H. Raffy, A.
Santander-Syro, K. Kadowaki, M. R. Norman, M. Randeria, and J. C. Campuzano,
Nondispersive Fermi arcs and the absence of charge ordering in the pseudogap phase
of Bi$_{2}$Sr$_{2}$CaCu$_{2}$O$_{8+\delta}$,
Phys. Rev. Lett. {\bf 96},107006 (2006).

\bibitem{McElroy06} K. McElroy, G.-H. Gweon, S. Y. Zhou, J. Graf, S. Uchida, H.
Eisaki, H. Takagi, T. Sasagawa, D.-H. Lee, and A. Lanzara,
Elastic scattering susceptibility of the high temperature superconductor
Bi$_{2}$Sr$_{2}$CaCu$_{2}$O$_{8+\delta}$:
a comparison between real and momentum space photoemission spectroscopies,
Phys. Rev. Lett. {\bf 96}, 067005 (2006).

\bibitem{Chatterjee07} U. Chatterjee, M. Shi, A. Kaminski, A. Kanigel, H. M.
Fretwell, K. Terashima, T. Takahashi, S. Rosenkranz, Z. Z. Li, H. Raffy, A.
Santander-Syro, K. Kadowaki, M. Randeria, M. R. Norman, and J. C. Campuzano,
Anomalous dispersion in the autocorrelation of angle-resolved photoemission
spectra of high-temperature Bi$_{2}$Sr$_{2}$CaCu$_{2}$O$_{8+\delta}$
superconductors, Phys. Rev. B {\bf 76}, 012504 (2007).

\bibitem{Restrepo23} F. Restrepo, J. Zhao, J. C. Campuzano, and U. Chatterjee,
Temperature and carrier concentration dependence of Fermi arcs in moderately
underdoped Bi$_{2}$Sr$_{2}$CaCu$_{2}$O$_{8+\delta}$ cuprate high-temperature
superconductors: A joint density of states perspective,
Phys. Rev. B {\bf 107}, 174519 (2023).

\bibitem{He14} Y. He, Y. Yin, M. Zech, A. Soumyanarayanan, M. M. Yee, T. Williams, M.
C. Boyer, K. Chatterjee, W. D. Wise, I. Zeljkovic, T. Kondo, T. Takeuchi, H. Ikuta,
P. Mistark, R. S. Markiewicz, A. Bansil, S. Sachdev, E. W. Hudson, and J. E. Hoffman,
Fermi surface and pseudogap evolution in a cuprate superconductor,
Science {\bf 344}, 608 (2014).

\bibitem{Norman98} M. R. Norman, H. Ding, M. Randeria, J. C. Campuzano, T. Yokoya,
T. Takeuchi, T. Takahashi, T. Mochiku, K. Kadowaki, P. Guptasarma, and D. G. Hinks,
Destruction of the Fermi surface in underdoped high-$T_{c}$ superconductors,
Nature {\bf 392}, 157 (1998).

\bibitem{Shi08} M. Shi, J. Chang, S. Pailh\'es, M. R. Norman, J. C. Campuzano, M.
M\'ansson, T. Claesson, O. Tjernberg, A. Bendounan, L. Patthey, N. Momono, M. Oda,
M. Ido, C. Mudry, and J. Mesot,
Coherent $d$-wave superconducting gap in underdoped La$_{2-x}$Sr$_{x}$CuO$_{4}$ by
angle-resolved photoemission spectroscopy,
Phys. Rev. Lett. {\bf 101}, 047002 (2008).

\bibitem{Sassa11} Y. Sassa, M. Radovi\'c, M. M\'ansson, E. Razzoli, X. Y. Cui, S.
Pailh\'es, S. Guerrero, M. Shi, P. R. Willmott, F. Miletto Granozio, J. Mesot, M. R.
Norman, and L. Patthey,
Ortho-II band folding in YBa$_{2}$Cu$_{3}$O$_{7-\delta}$ films revealed by
angle-resolved photoemission,
Phys. Rev. B {\bf 83}, 140511(R) (2011).

\bibitem{Comin14} R. Comin, A. Frano, M. M. Yee, Y. Yoshida, H. Eisaki, E. Schierle,
E. Weschke, R. Sutarto, F. He, A. Soumyanarayanan, Yang He, M. L. Tacon, I. S.
Elfimov, Jennifer E. Hoffman, G. A. Sawatzky, B. Keimer, and A. Damascelli,
Charger order driven by Fermi-arc instability in
Bi$_{2}$Sr$_{2-x}$La$_{x}$CuO$_{6+\delta}$,
Science {\bf 343}, 390 (2014).

\bibitem{Kaminski15} A. Kaminski, T. Kondo, T. Takeuchi, and G. Gu, Pairing,
pseudogap and Fermi arcs in cuprates, Phil. Mag. {\bf 95}, 453 (2015).

\bibitem{Zeng22} M. Zeng, X. Li, Y. Wang, and S. Feng, Influence of impurities on
the electronic structure in cuprate superconductors, Phys. Rev. B {\bf 106},
054512 (2022); Y. Liu, Y. Lan, and S. Feng, Peak structure in the self-energy of
cuprate superconductors, Phys. Rev. B {\bf 103}, 024525 (2021).

\bibitem{Anderson87a} P. W. Anderson, 50 years of the Mott phenomenon: insulators,
magnets, solids and superconductors as aspects of strong-repulsion theory, in
{\it Frontiers and Borderlines in Many Particle Physics}, edited by R. A. Broglia
and J. R. Schrieffer (North-Holland, Amsterdam, 1987), p. 1.

\bibitem{Zhang88} F.C. Zhang and T.M. Rice, Effective Hamiltonian for the
superconducting Cu oxides, Phys. Rev. B {\bf 37}, 3759(R) (1988).

\bibitem {Lee06} See, e.g., the review, P. A. Lee, N. Nagaosa, and X.-G. Wen,
Doping a Mott insulator: Physics of high-temperature superconductivity, Rev. Mod.
Phys. {\bf 78}, 17 (2006).

\bibitem{Edegger07} See, e.g., the review, B. Edegger, V. N. Muthukumar, and C.
Gros, Gutzwiller-RVB theory of high-temperature superconductivity: Results from
renormalized mean-field theory and variational Monte Carlo calculations, Adv.
Phys. {\bf 56}, 927 (2007).

\bibitem {Spalek22} J. Spalek, M. Fidrysiak, M. Zegrodnik, and A. Biborski,
Superconductivity in high-$T_c$ and related strongly correlated systems from
variational perspective: Beyond mean field theory, Phys. Rep. {\bf 959}, 1 (2022).

\bibitem {Feng0494} S. Feng, J. Qin, and T. Ma, A gauge invariant dressed holon and
spinon description of the normal state of underdoped cuprates,
J. Phys.: Condens. Matter {\bf 16},343 (2004); S. Feng, Z. B. Su, and L. Yu,
Fermion-spin transformation to implement the charge-spin separation,
Phys. Rev. B {\bf 49}, 2368 (1994).

\bibitem{Feng15} See, e.g., the review, S. Feng, Y. Lan, H. Zhao, L. Kuang, L. Qin,
and X. Ma, Kinetic-energy driven superconductivity in cuprate superconductors, Int.
J. Mod. Phys. B {\bf 29}, 1530009 (2015).

\bibitem{Feng0306} S. Feng, Kinetic energy driven superconductivity in doped
cuprates, Phys. Rev. B {\bf 68}, 184501 (2003); S. Feng, T. Ma, and H. Guo,
Magnetic nature of superconductivity in doped cuprates, Physica C {\bf 436},
14 (2006).

\bibitem{Feng12} S. Feng, H. Zhao, and Z. Huang,
Two gaps with one energy scale in cuprate superconductors, Phys. Rev. B. {\bf 85},
054509 (2012); Phys. Rev. B {\bf 85}, 099902(E) (2012).

\bibitem{Feng15a} S. Feng, L. Kuang, and H. Zhao,
Electronic structure of cuprate superconductors in a full charge-spin recombination
scheme, Physica C {\bf 517}, 5 (2015).

\bibitem{Schrieffer95} J. R. Schrieffer,
Ward's identity and the suppression of spin fluctuation superconductivity, J. Low
Temp. Phys. {\bf 99}, 397 (1995).

\bibitem{Xu23a} K.-J. Xu, Q. Guo, M. Hashimoto, Z.-X. Li, S.-D. Chen, J. He, Y. He,
C. Li, M. H. Berntsen, C. R. Rotundu, Y. S. Lee, T. P. Devereaux, A. Rydh, D.-H. Lu,
D.-H. Lee, O. Tjernberg, Z.-X. Shen, Bogoliubov quasiparticle on the gossamer Fermi
surface in electron-doped cuprates, Nat. Phys. 19, 1834 (2023).

\bibitem{Gao19} D. Gao, Y. Mou, Y. Liu, S. Tan, and S. Feng,
Autocorrelation of quasiparticle spectral intensities and its connection with
quasiparticle scattering interference in cuprate superconductors,
Phil. Mag. {\bf 99}, 752 (2019).

\bibitem{Eisaki04} H. Eisaki, N. Kaneko, D. L. Feng, A. Damascelli, P. K. Mang, K.
M. Shen, Z.-X. Shen, and M. Greven,
Effect of chemical inhomogeneity in bismuth-based copper oxide superconductors,
Phys. Rev. B {\bf 69}, 064512 (2004).

\bibitem{Fujita05} K. Fujita, T. Noda, K. M. Kojima, H. Eisaki, and S. Uchida,
Effect of disorder outside the CuO$_{2}$ planes on $T_{c}$ of copper oxide
superconductors, Phys. Rev. Lett. {\bf 95}, 097006 (2005).

\bibitem{McElroy05} K. McElroy, Jinho Lee, J. A. Slezak, D.-H. Lee, H. Eisaki, S.
Uchida, and J. C. Davis,
Atomic-scale sources and mechanism of nanoscale electronic disorder in
Bi$_{2}$Sr$_{2}$CaCu$_{2}$O$_{8+\delta}$, Science {\bf 309}, 1048 (2005).

\bibitem{Kondo07} T. Kondo, T. Takeuchi, A. Kaminski, S. Tsuda, and S. Shin,
Evidence for two energy scales in the superconducting state of optimally doped
(Bi,Pb)$_{2}$(Sr,La)$_{2}$CuO$_{6+\delta}$,
Phys. Rev. Lett. {\bf 98}, 267004 (2007).

\bibitem{Hashimoto08} M. Hashimoto, T. Yoshida, A. Fujimori, D. H. Lu, Z.-X. Shen,
M. Kubota, K. Ono, M. Ishikado, K. Fujita, and S. Uchida,
Effects of out-of-plane disorder on the nodal quasiparticle and superconducting gap
in single-layer Bi$_{2}$Sr$_{1.6}$L$_{0.4}$CuO$_{6+\delta}$(L=La,Nd,Gd),
Phys. Rev. B {\bf 79},144517 (2009).

\bibitem{Zhu04} L. Zhu, P. J. Hirschfeld, and D. J. Scalapino,
Elastic forward scattering in the cuprate superconducting state, Phys. Rev.
B. {\bf 70}, 214503 (2004).

\bibitem{Nunner06} T. S. Nunner, W. Chen, B. M. Andersen, A. Melikyan, and P. J.
Hirschfeld,
Fourier transform spectroscopy of $d$-wave quasiparticles in the presence of
atomic scale pairing disorder, Phys. Rev. B {\bf 73}, 104511 (2006).

\bibitem{JXZhu2000} J.-X. Zhu, C. S. Ting, and C.-R. Hu,
Effect of unitary impurities on non-STM types of tunneling in high-$T_{c}$
superconductors, Phys. Rev. B {\bf 62}, 6027 (2000).

\bibitem{Martin2002} I. Martin, A. V. Balatsky, and J. Zaanen,
Impurity states and interlayer tunneling in high temperature superconductors,
Phys. Rev. Lett. {\bf 88}, 097003 (2002).

\bibitem{Zhu03a} L. Zhu, W. A. Atkinson, and P. J. Hirschfeld,
Two impurities in a $d$-wave superconductor: Local density of states, Phys. Rev.
B {\bf 67}, 094508 (2003).

\bibitem{Atkinson03a} W. A. Atkinson, P. J. Hirschfeld, and L. Zhu,
Quantum interference in nested $d$-wave superconductors: A real-space perspective,
Phys. Rev. B {\bf 68}, 054501 (2003).

\bibitem{Atkinson00a} W. A. Atkinson, P. J. Hirschfeld, A. H. MacDonald, and K.
Ziegler, Details of disorder matter in 2D $d$-Wave superconductors, Phys. Rev.
Lett. {\bf 85}, 3926 (2000).


\end{thebibliography}
\end{document}